\documentclass[preprint,12pt]{elsarticle}

\usepackage{amssymb}
\usepackage{amsthm}

\usepackage{amsmath,amstext,amsfonts}

\usepackage{booktabs}  
\usepackage{colortbl}  
\usepackage{xcolor}  

\usepackage{caption}
\usepackage{subcaption}

\DeclareMathOperator\erfinv{erfinv}

\graphicspath{{./figures/}}

\journal{}

\begin{document}

\begin{frontmatter}


\author{Benjamin B. Schroeder}
\ead{benjamin.schroeder@utah.edu}
\author{Sean T. Smith}
\author{Philip J. Smith}
\address{Department of Chemical Engineering\\
University of Utah\\
Salt Lake City, UT 84112}

\author{Thomas H. Fletcher}
\address{Department of Chemical Engineering\\
Brigham Young University\\
Provo, UT 84602}

\author{Andrew Packard, Michael Frenklach, Arun Hegde, Wenyu Li, James Oreluk}
\address{Department of Mechanical Engineering\\
University of California, Berkeley\\
Berkeley, CA 94720}

\title{Scale-Bridging Model Development for Coal Particle Devolatilization}


\author{}

\address{}

\begin{abstract}
When performing large-scale, high-performance computations of multi-physics applications, it is common to limit the complexity of physics sub-models comprising the simulation.
For a hierarchical system of coal boiler simulations a scale-bridging model is constructed to capture characteristics appropriate for the application-scale from a detailed coal devolatilization model.
Such scale-bridging allows full descriptions of scale-applicable physics, while functioning at reasonable computational costs.
This study presents a variation on multi-fidelity modeling with a detailed physics model, the chemical percolation devolatilization model, being used to calibrate a scale-briding model for the application of interest.
The application space provides essential context for designing the scale-bridging model by defining scales, determining requirements and weighting desired characteristics.
A single kinetic reaction equation with functional yield model and distributed activation energy is implemented to act as the scale-bridging model-form.
Consistency constraints are used to locate regions of the scale-bridging model's parameter-space that are consistent with the uncertainty identified within the detailed model.
Ultimately, the performance of the scale-bridging model with consistent parameter-sets was assessed against desired characteristics of the detailed model and found to perform satisfactorily in capturing thermodynamic trends and kinetic timescales for the desired application-scale.
Framing the process of model-form selection within the context of calibration and uncertainty quantification allows the credibility of the model to be established.
\end{abstract}

\begin{keyword}
Scale-Bridging Model \sep Uncertainty \sep Devolatilization \sep Reduced Order Modeling



\end{keyword}

\end{frontmatter}


\section{Introduction}
\label{sec:intro}

A typical approach to multi-fidelity modeling is to identify the discrepancy between model fidelities when simulations are run with the same inputs, and then use the lower fidelity model with the discrepancy to make predictions\cite{Alexandrov2001,Wang2015}.
Instead of that approach, within this application a scale-bridging model (SBM) will be calibrated to a more detailed physics model and then used for prediction.
Scale-bridging is a technique commonly found in simulation science \cite{vanderHoef2006,Marchisio2007}, where sub-model complexity is limited by the simulation's resolution.
Scales herein refer to temporal and/or spatial.
The term model-form uncertainty will be used herein to refer to differences between a model prediction and reality.
This idea was also referred to as model inadequacy within the seminal paper upon this topic by Kennedy and O'Hagan (2001) \cite{Kennedy2001}.

Large-scale multi-physics based simulations often try to capture physical phenomena for which detailed physics models exist.
Such detailed physics models often function at scales smaller than the simulations operate.
While the detailed models may predict the exact physics required, the simulation cannot economically function at such small-scales.
Where models acting at small-scales captured more detailed descriptions of physical phenomena occurring, only attributes of those physics that impact the large-scale application are needed for the simulations.
In order to capture physics of small-scale models in large-scale simulations, SBMs are employed.
SBMs are formulated to capture characteristics of the detailed model for conditions and scales appropriate to the application.
While calibrating and uncertainty quantification techniques are not required for SBM formulation, it is through such processes that credible models can be developed.
Beyond simply capturing physical traits at the appropriate scale and calibrating parameters, credibility is also desired in order to provide confidence in exercising the SBM.

First and foremost the SBM must be able to produce results within the uncertainty of the more detailed model's results.
Ideally, the SBM will  capture the full range of the detailed model's uncertainty, thereby allowing the full impact of the uncertainty in this set of physics be explored within the simulation applications.
A SBM is based upon fewer parameters than the progenitor detailed model, and thus should allow easier propagation of uncertainty into the ultimate applications.
That being said, sensitivity to the SBM's parameters found through application simulations will be difficult to map back to parameters in the more detailed model because the fitted SBM parameters may not have true physical interpretations.
Emulators could be created to capture the desired physics behaviors with cheap function evaluations, but the physic basis of SBM are believed to provide greater credibility for extrapolations.

\begin{figure}[t!]
\centering
\includegraphics[width=.9\columnwidth,keepaspectratio]{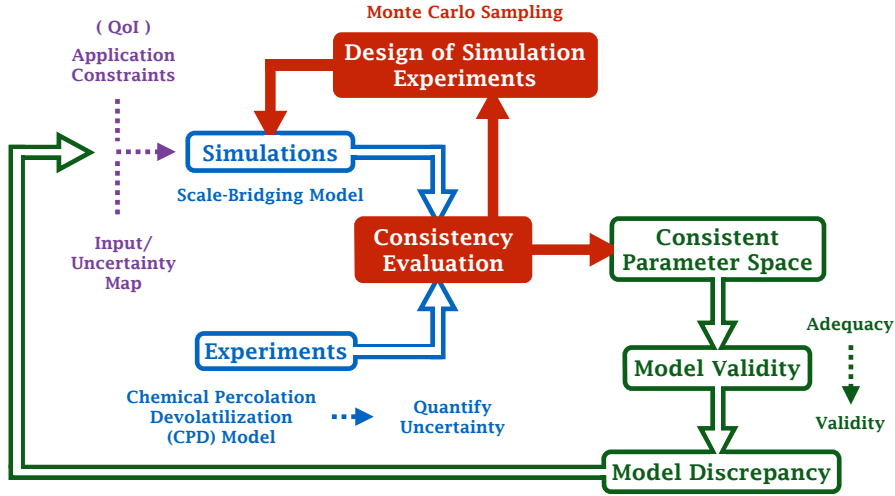}
\caption{Diagram of the method used to develop a scale-bridging coal devolatilization model}
\label{fig:Model_Proc_Flow}
\end{figure}

A diagram of the approach taken herein for credible model development is shown in Figure~\ref{fig:Model_Proc_Flow}.
Initially, the model will be tested for basic adequacy.
In the terms of the flow diagram, adequacy is assessed through the comparison of simulation outputs with experimental data.
Such comparisons are restricted to characteristics that can be quantified into comparable terms.
While it is more typical to compare a model with experimental data to infer the model-form error, an activity commonly known as model validation, a detailed model will be used instead for this scale-bridging model development process.
Consistency evaluations determine if a subset of the parameter-space exists that allows the model to meet constraints specified for adequacy.
Assuming that the model is shown adequate for the application, analysis of the model-form error, or discrepancy between SBM and corresponding detailed model outputs, will drive the model development towards credibility.
While this document focuses upon a specific application, the general concept of utilizing SBMs to capture scale-appropriate characteristics from detailed physical models, with quantified uncertainty, is applicable to a wide array of engineering problems.

The diagram also provides an overview of this document's organization.  First, the application of interest is presented.
Next, the detailed physics model is introduced and its characteristics explored to determine which features are necessary for the simulation-scale.
Input sensitivity and uncertainty quantification are also analyzed for the detailed model at this point.
Following this will be a summary of the scale-bridging model-form development.
The methodology used for consistency testing will then be defined as well as the application-based constraints upon which consistency is evaluated.
Once consistent regions of the parameter-space have been found, the performance metrics of the current model-form are assessed and refinements to the model-form are proposed based upon the model discrepancy.

\section{Application Space}
\label{sec:application_space}

A suite of large-scale computational fluid dynamics (CFD) simulations of various scales of industrially-relevant coal boiler configurations will be the application space for a SBM.
The boilers' scales span from tens of MWe up to 500 MWe, with an emphasis on larger-scales due to these being most informative for industrial applications.
This suite of simulations will be continually refined over a time frame of years and each year a validation and uncertainty quantification (VUQ) exploration of each boiler configuration will be conducted to appraise its current status.
High-performance computer-simulations take significant computational resources, so simulation runs are limited.
Thus, each VUQ analysis will be confined to exploring two/three uncertain parameters per year.
Each simulation contains hundreds of uncertain parameters and tens of uncertain physics models.
Focusing on two or three parameters from this list drastically limits what can be studied within a series of simulation runs.

For year one the piece of physics considered to have uncertainty to which the simulation outputs of interest will be most sensitive was coal devolatilization.
Thus, coal devolatilization physics will provide a parameter for the first year's VUQ study.
When studying coal devolatilization in this context, the application space cannot be ignored.
Instantaneous function evaluations are necessary for the large-scale simulation, where devolatilization is calculated for each grid-point at each time-step.
Devolatilization is the mechanism by which non-oxidized gases and tars move from the solid to gaseous phase for coal combustion or other thermal treatment, and thus have a significant influence upon the entire simulation.
Detailed models of coal devolatilization exist \cite{Solomon1988,Niksa1991,Fletcher1992} and have been successful in describing experimental data, but are computationally too expensive to incorporate into large-scale CFD simulations.
Simply put, a function with cheap evaluations that accurately captures the physical process was needed, or a SBM.
The SBM needs to function within the CFD's limitations, while still capturing the necessary physical characteristics.

This devolatilization model is being developed into Arches - an open-source large-eddy simulation package for multi-phase, turbulent, reacting flow on massively parallel architectures (up to 250,000 cores per simulation) \cite{Spinti2008,Pedel2014,Smith2014}.
Arches has been previously used to simulate industrial flares, once-through steam generators, pool fires, and boilers.
The study for this manuscript is limited to coal applications - and the subset of interest is coal pyrolysis/devolatilization.
As a point of reference, Arches has always modeled the coal devolatilization process using classical two-step mechanisms \cite{Kobayashi1977,Ubhayakar1977} or a one-step mechanism that is fit to CPD with no parameter uncertainty \cite{Yamamoto2011}.

Devolatilization is a chemical process that can be viewed from a kinetic or thermodynamic perspective.
Determination of the effects and importance of attributes of the physical process, as they pertain to the scale at which the application occurs, must be investigated.
Such a probe can be undertaken with a detailed model that functions at small-scales.

\section{The CPD Model}
\label{sec:cpd_model}

The detailed physics description of coal devolatilization used is the chemical percolation devolatilization (CPD) model developed by Fletcher et al. (1992) \cite{Fletcher1992}.
CPD is an example of a devolatilization model that produces accurate results for small temporal-scale resolution through relatively expensive evaluations.
Within CPD, NMR spectroscopic data is used to characterize the composition of a specified coal type.
The CPD model includes a bridge-breaking reaction scheme, lattice statistics, percolation theory, and chemical phase-equilibrium calculations.
CPD assumes a uniform temperature throughout the particle, or extremely small thermal Biot number, allowing the model outputs to be scaled by the mass of the desired particle size.
In collaboration with Professor Fletcher of Brigham Young University, a MATLAB version of the CPD code \cite{CPD} produces the detailed physics data for this study.
Professor Fletcher provided uncertainty ranges for 13 uncertain model parameters contained within CPD, as shown in Table~\ref{tab:CPD_uncertain}.
\begin{table}[t!]
\caption{CPD uncertain parameters and uncertainty ranges solicited from Professor Fletcher of Brigham Young University}
\label{tab:CPD_uncertain}
\centering
\resizebox{4in}{!}{
\begin{tabular}{c|c|c|cc}
\toprule
\rowcolor{black!20} Parameter & Nominal & Uncertainty & Max. & Min. \\
\midrule
\rowcolor{black!10} \multicolumn{5}{c}{Coal Specific (Utah Sufco bituminous)} \\
p$_0$ [-] & 0.483 & 0.03 & 0.513 & 0.453  \\
c$_0$ [-] & 0.0827 & - & - & - \\
$\sigma$ + 1 [-] & 4.78 & 0.2 & 4.98 & 4.58 \\
M$_{clust}$ [kg/kmol] & 457.8 & 20 & 477.8 & 437.8 \\
m$_\delta$ [kg/kmol] & 45.7 & 2 & 47.7 & 43.7 \\
\rowcolor{black!10} \multicolumn{5}{c}{General CPD Model} \\
A$_b$ [s$^{-1}$] & $2.6E{+15}$ & $5\%$ & $2.73E{+15}$ & $2.47E{+15}$ \\
E$_b$ [cal/mol] & 55,400 & $5\%$ & 58,170 & 52,630 \\
$\sigma_b$ [cal/mol] & 1,800 & $5\%$ & 1,890 & 1,710 \\
ac [-] & 0.9 & $0.05\%$ & 0.90045 & 0.89955 \\
ec [-] & 0 & - & - & - \\
A$_g$ [s$^{-1}$] & $3.0E{+15}$ & $5\%$ & $3.15E{+15}$ & $2.85E{+15}$ \\
E$_g$ [cal/mol] & 69,000 & $5\%$ & 72,450 & 65,550 \\
$\sigma_g$ [cal/mol] & 8,100 & $5\%$ & 8,505 & 7,695 \\
A$_{cr}$ [s$^{-1}$] & $3.0E{+15}$ & $5\%$ & $3.15E{+15}$ & $2.85E{+15}$ \\
E$_{cr}$ [cal/mol] & 65,000 & $5\%$ & 68,250 & 61,750 \\
\bottomrule
\end{tabular}}
\end{table}
These estimated ranges can be treated as uncertainty intervals based upon expert opinion.
A secondary aspect of utilizing CPD for this study is that the credibility of the SBM will be leveraging CPD's credibility.

\subsection{Temperature and Heating-Rate Effects}
\label{sec:heating_temp_effect}
An exploratory investigation of CPD was undertaken in order to identify scale-appropriate attributes that the SBM should aim to capture.
When developing models of coal devolatilization physics, two system conditions are typically considered to be process controlling: the rate at which coal is heated and the ultimate temperature reached (hold-temperature).
Due to the limitations of the ultimate application scale for the devolatilization model, a simplistic SBM was desired.
Thus, the potential of eliminating one of these controlling system conditions from the SBM model-form was considered.
The initial investigation of the heating-rate and hold-temperature effects spanned a wider range of system conditions than was anticipated to be relevant for the application.
Nevertheless, this analysis was used to gain a wider grasp of potential implications.
Devolatilization is sensitive to coal composition, but a single coal type, Utah Sufco bituminous, was used throughout the current analysis.
Pressure can also affect this phenomena \cite{Fletcher1992}, but atmospheric pressure was assumed throughout this analysis because the boilers under consideration operate near atmospheric pressure.
Nominal CPD parameter values were used through the hold-temperature and heating-rate analysis.

\subsubsection{Hold-Temperature}
\label{subsec:hold_temp}

\begin{figure}[t!]
\centering
\includegraphics[width=0.6\columnwidth,keepaspectratio]{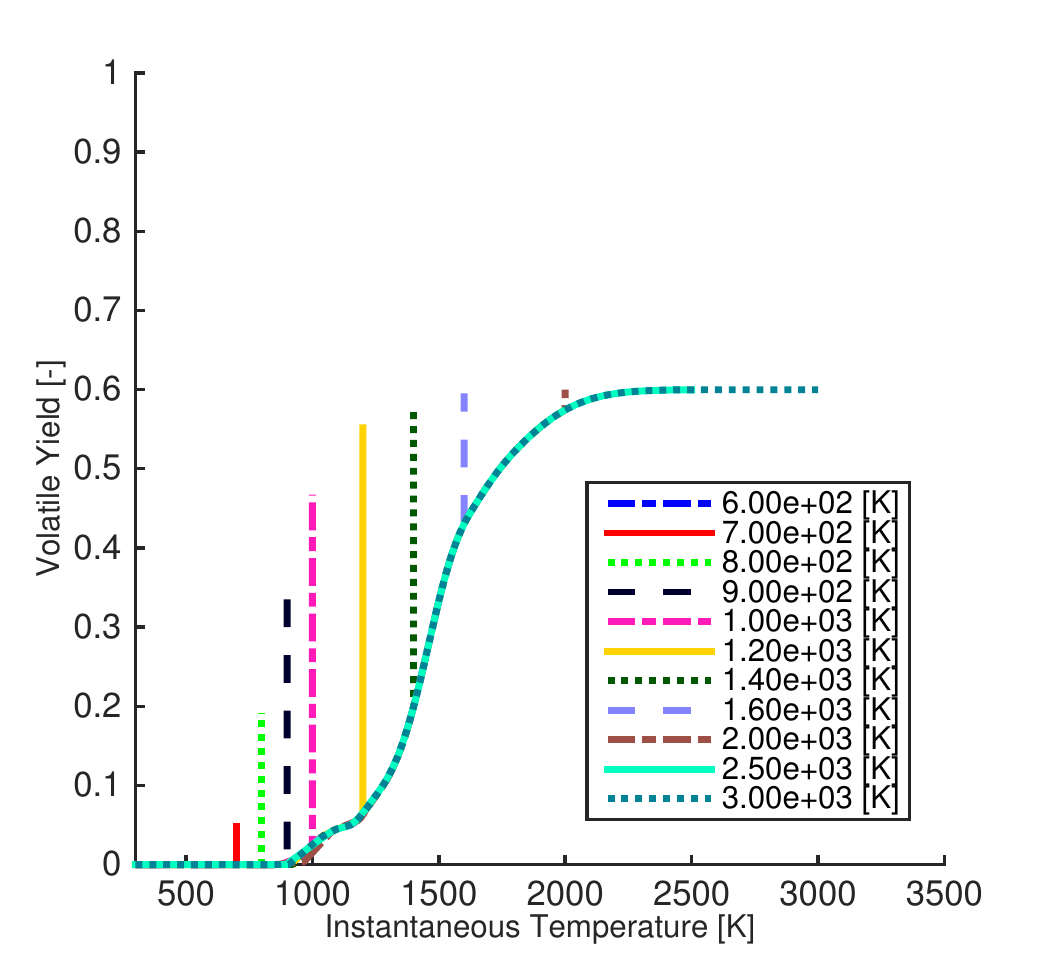}
\caption{Volatile yield traces of Utah Sufco bituminous coal devolatilization for a range of hold-temperatures over their instantaneous temperature.
To create the trace profiles, the coal was linearly heated from 300 K to the specified hold-temperature at a rate of 1E9 K/s and then held at the hold-temperature for ten seconds.
Nominal parameter values from Table~\ref{tab:CPD_uncertain} were used.}
\label{fig:ConstRtraces}
\end{figure}

To investigate the impact that the ultimate hold-temperature had upon coal's devolatilization, CPD calculations were made over a range of hold-temperatures spanning 500 K to 3,500 K as is shown by the family of volatile yield traces in Figure~\ref{fig:ConstRtraces}.
These calculations all used a linear heat-up rate of 1E9 K/s, which was assumed to effectively represent instantaneous heating to the hold-temperature.
Instantaneous heating of the coal was desired in order to isolate the hold-temperature's effect.
The initial coal temperature was specified as 300 K.
Once the hold-temperature was reached, the coal was held at that temperature for ten seconds.
This hold-time was assumed to be the infinite time-scale for the system of interest.
For applications where lower temperatures and heating-rates are prevalent, such an assumption would not be appropriate.

\begin{figure}[t!]
\centering
\includegraphics[width=0.6\columnwidth,keepaspectratio]{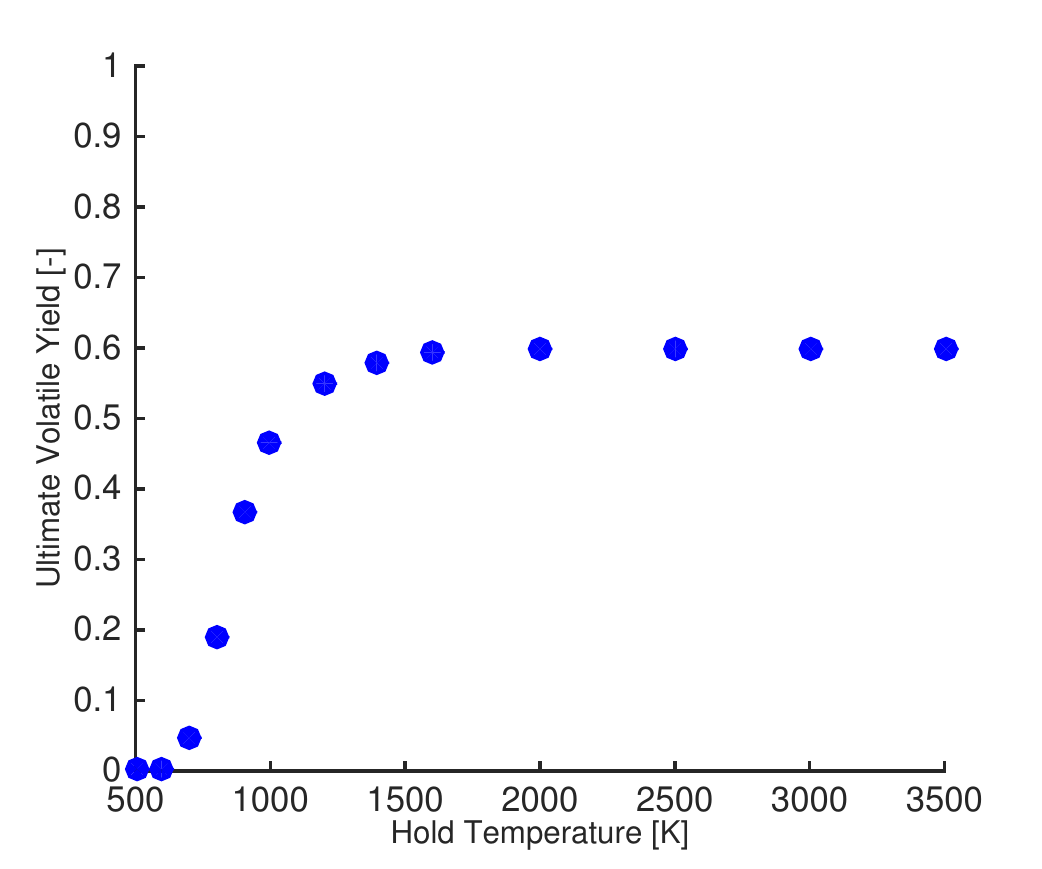}
\caption{Ultimate volatile yield for Utah Sufco bituminous coal due to devolatilization over a range of hold-temperatures.
Data points were extracted from traces shown in Figure~\ref{fig:ConstRtraces} and were the product of linearly heating the coal from 300 K to the specified hold-temperature at a rate of 1E9 K/s and holding the coal at the hold-temperature for ten seconds.
Nominal parameter values from Table~\ref{tab:CPD_uncertain} were used.}
\label{fig:ConstRUltYield}
\end{figure}

Vertical spikes in the volatile yield traces shown in Figure~\ref{fig:ConstRtraces} are due to the coal reaching its specified hold-temperature and then continuing to produce volatiles until reaching the effective equilibrium for that hold-temperature.
The resulting equilibrium curve can be visualize in Figure~\ref{fig:ConstRUltYield}, where data-points shown are the last points of the volatile yield traces in Figure~\ref{fig:ConstRtraces}.

It is evident that the ultimate volatile yield is strongly affected by the hold-temperature.
The ultimate volatile yield curve represents an equilibrium curve for the application space of interest, but likely would change for applications in other domains such as underground in situ heating of coal where the heating-rates are much slower and the timescales are far longer.
For the current application it appears that devolatilization initiates around 600 K and  minimal changes occur above 1,600 K.
While the asymptotic behavior for high temperatures can be debated, this is not considered presently.
Another use for this curve could be to act as a yield model.
Volatile yield traces for temperatures above 2,500 K were found to vary minimally, as can be noted in the overlap of traces for 2,500 K and 3,000 K, and thus were not included in Figure~\ref{fig:ConstRtraces}.

\begin{figure}[t!]
\centering
\includegraphics[width=0.6\columnwidth,keepaspectratio]{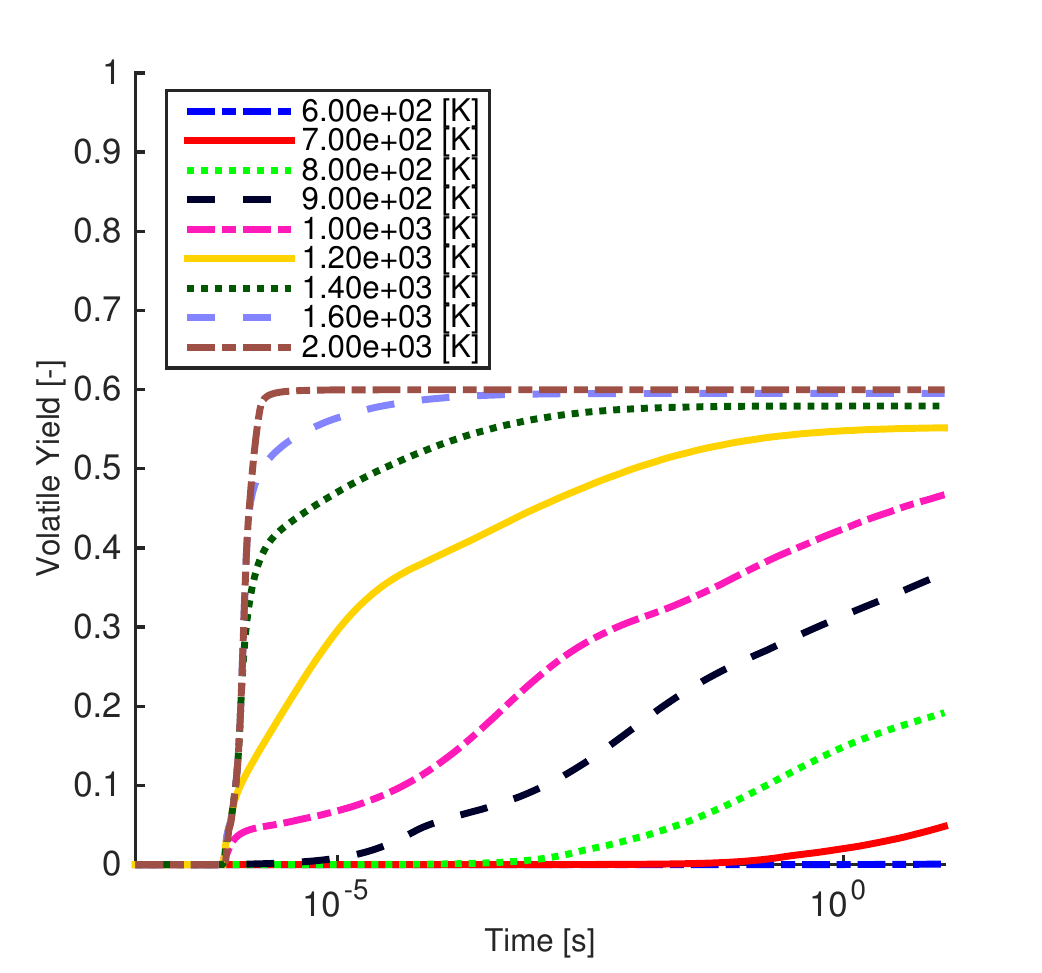}
\caption{Volatile yield traces for Utah Sufco bituminous coal due to devolatilization for a range of hold-temperatures as they evolve over time.
Coal was linearly heated at a rate of 1E9 K/s from 300 K to the hold-temperature and held at that temperature for 10 seconds.
Nominal parameter values from Table~\ref{tab:CPD_uncertain} were used.}
\label{fig:ConstRtime}
\end{figure}

Volatile yield is also a function of time, with coal exposed to lower temperatures losing volatiles over a longer period of time.
This temporal functionality is shown in Figure~\ref{fig:ConstRtime}.
From the time traces it is clear that the chosen infinite time is in fact not infinite for coal at lower temperatures.
For coal that reaches temperatures above 1,200 K, this infinite time assumption appears sufficient.
Even with the now recognized deficiencies in this assumption, coal will not spend longer than ten seconds in the boiler applications of interest, which allows this assumption to be retained.
All of the temporal traces above 2,000 K overlap and were left off Figure~\ref{fig:ConstRtime} for visual clarity.

\subsubsection{Heating Rate}
\label{subsec:heating_rate}

\begin{figure}[t!]
\centering
\includegraphics[width=0.6\columnwidth,keepaspectratio]{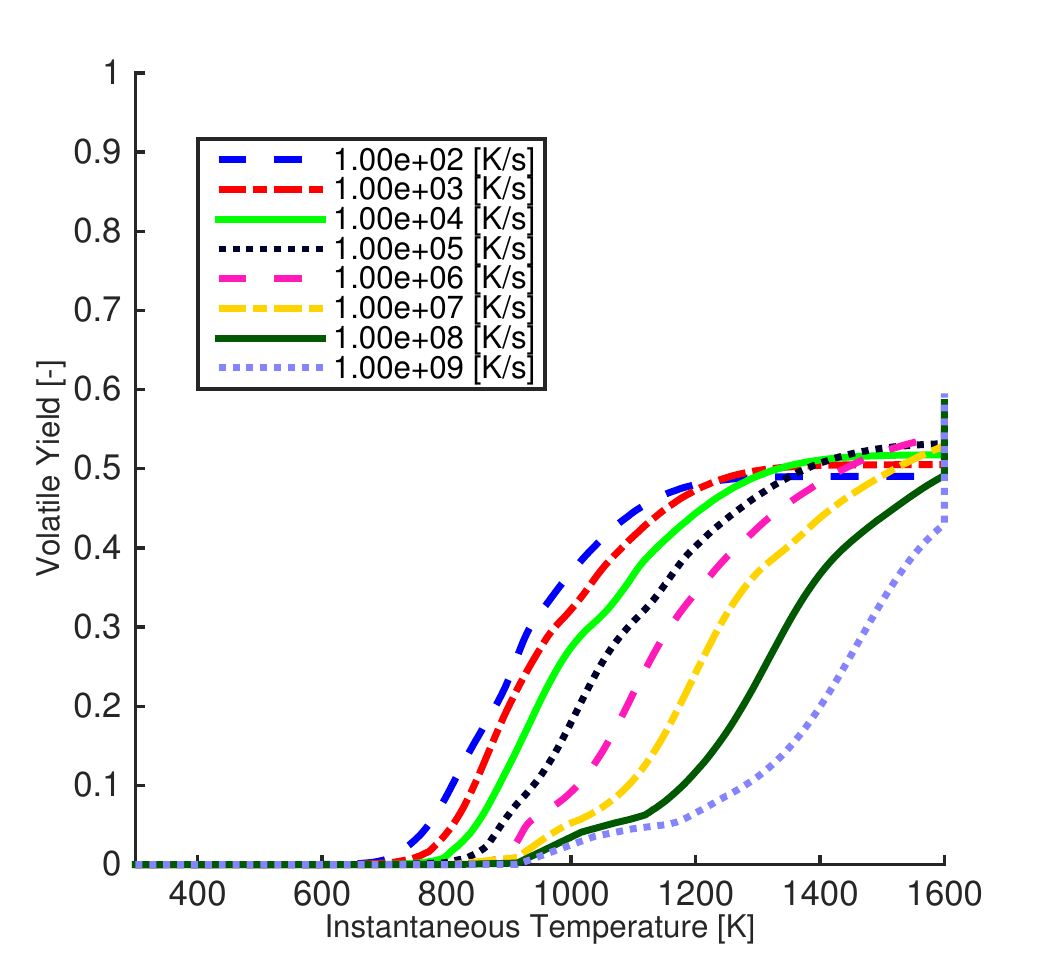}
\caption{Volatile yield traces for Utah Sufco bituminous coal due to devolatilization for a range of heat-up rates over their instantaneous temperatures.
These traces were calculated by CPD using a final hold-temperature of 1,600 K and were held at that hold-temperature for ten seconds.
For the range of heat-up rates shown, the coal was linearly heated from 300 K to the hold-temperature.
Nominal parameter values from Table~\ref{tab:CPD_uncertain} were used.}
\label{fig:ConstTtraces}
\end{figure}

The impact of the heating-rate on coal's volatile yield was examined in the same manner as the hold-temperature.
Heating-rates ranging from 1E2 to 1E9 K/s were examined, as shown in Figure~\ref{fig:ConstTtraces}.
To produce the volatile yield traces shown, a hold-temperature of 1,600 K was used, which was reached after linearly heating the coal from 300 K.
Again, a hold-time of ten seconds was assumed to represent infinite time.

\begin{figure}[t!]
\centering
\includegraphics[width=0.6\columnwidth,keepaspectratio]{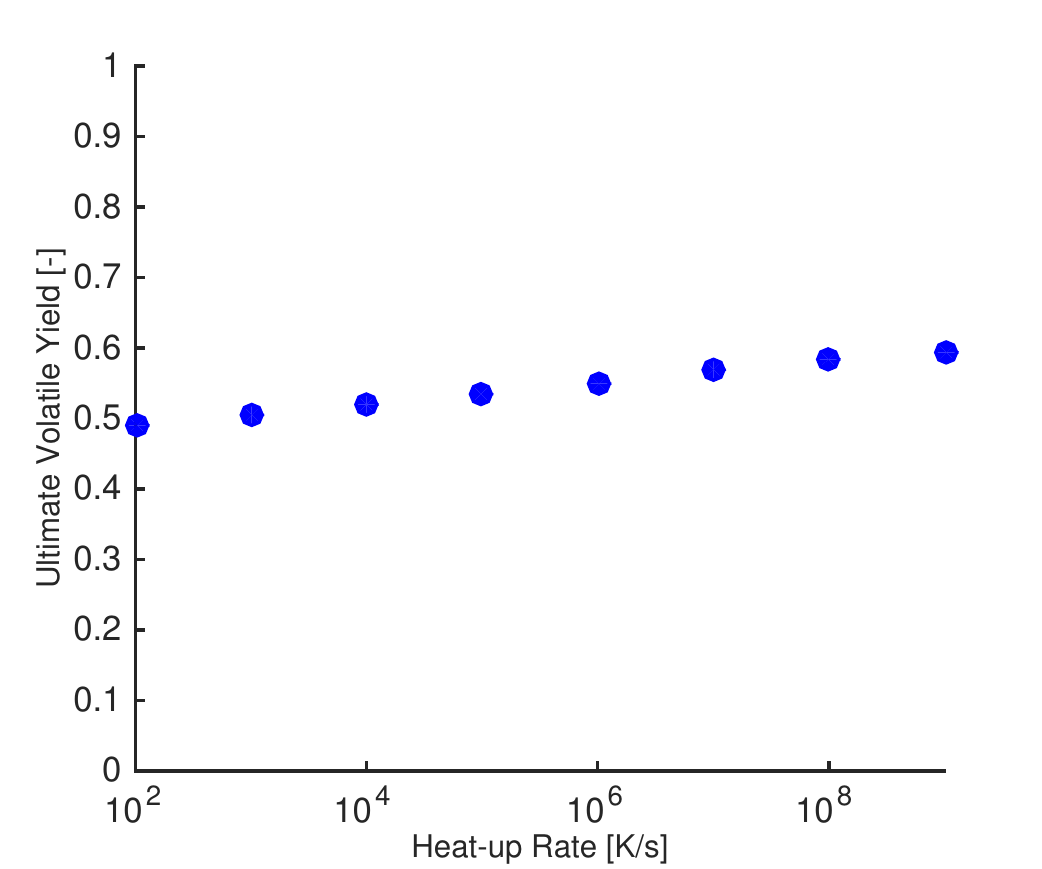}
\caption{Ultimate volatile yield for Utah Sufco bituminous coal due to devolatilization over a range of heat-up rates.
Data points were extracted from the data traces shown in Figure~\ref{fig:ConstTtraces} and were produced by linearly heating the coal at the specified rates from 300 K to 1,600 K and holding the coal at that temperature for ten seconds.
Nominal parameter values from Table~\ref{tab:CPD_uncertain} were used.}
\label{fig:ConstTUltYield}
\end{figure}

\begin{figure}[t!]
\centering
\includegraphics[width=0.6\columnwidth,keepaspectratio]{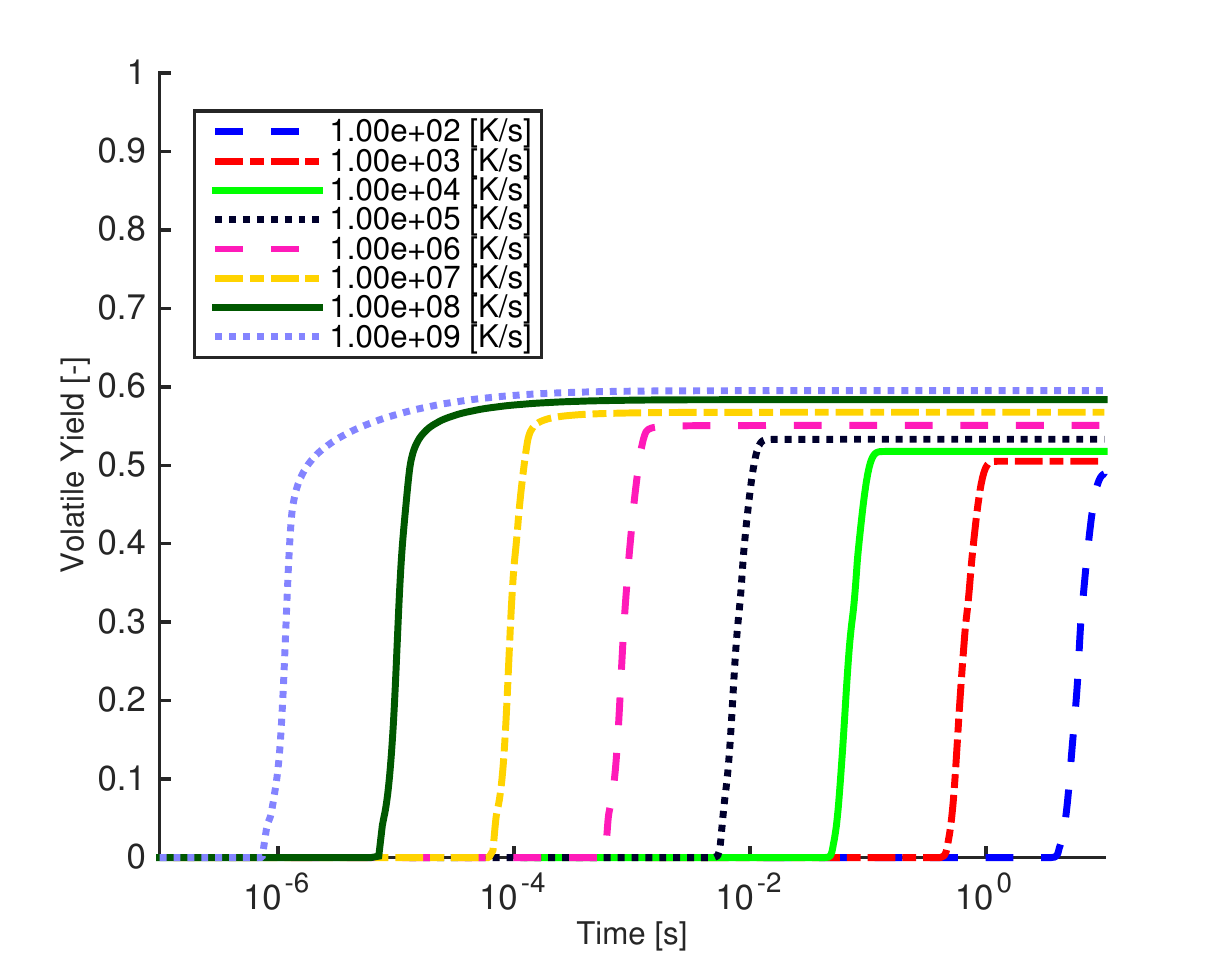}
\caption{Volatile yield traces for Utah Sufco bituminous coal due to devolatilization for a range of heating-rates as they evolve over time.
Traces were created by linearly heating the coal at the specified rates from 300 K to 1,600 K and holding the coal at that temperature for ten seconds.
Nominal parameter values from Table~\ref{tab:CPD_uncertain} were used.}
\label{fig:ConstTtime}
\end{figure}

Heating-rates evidently strongly affect devolatilization kinetics.
The traces for different heating-rates have different instantaneous temperatures for equivalent volatile yields, i.e., for coal heated at 1E9 K/s the instantaneous temperature for $20 \%$ volatile yield occurs around 1,400 K, while for coal heated at 1E2 K/s this occurs near 850 K.
While the heating-rates demonstrate strong effects upon the kinetics, the effect upon the equilibrium curve shown in Figure~\ref{fig:ConstTUltYield} is less significant than was noted for the hold-temperature in Figure~\ref{fig:ConstRUltYield}.
To verify that the heating-rates are functioning as expected, Figure~\ref{fig:ConstTtime} was created to ensure the heating of the coal had approximately order of magnitude spacing.


\subsection{Uncertainty Quantification}
\label{subsec:UQ}
In order to ensure that behaviors across the conditions of interest were sufficiently incorporated into the SBM development, a design of experiments (DOE) was specified.
The design was meant to cover regions of interest where CPD was believed to be accurate and/or which were important for the application boiler simulations.
A grid-wise DOE covering five hold-temperatures and three heating-rates, as shown in Table~\ref{tab:doetable}, was created to meet this objective.
\begin{table}[t!]
\caption{Matrix of 15 design of experiments conditions}
\label{tab:doetable}
\centering
\resizebox{0.6\columnwidth}{!}{
\begin{tabular}{p{2.2cm} | p{2.2cm} | p{2.2cm}}
\toprule
\rowcolor{black!20} \multicolumn{3}{c}{Hold-Temperatures [K] : Heating Rates [K/s]} \\
\midrule
\leavevmode\hphantom{1,}700 : 1E4 & \leavevmode\hphantom{1,}700 : 1E5 & \leavevmode\hphantom{1,}700 : 1E6 \\
1,000 : 1E4 & 1,000 : 1E5 & 1,000 : 1E6 \\
1,300 : 1E4 & 1,300 : 1E5 & 1,300 : 1E6 \\
1,600 : 1E4 & 1,600 : 1E5 & 1,600 : 1E6 \\
2,400 : 1E4 & 2,400 : 1E5 & 2,400 : 1E6 \\
\bottomrule
\end{tabular}}
\end{table}
Considering the variance that can be noted in the family of volatile yield traces shown in Sec.~\ref{sec:heating_temp_effect}, the SBM will be challenged to capture such dynamic characteristics.

\begin{figure}[t!]
\centering
\includegraphics[width=0.9\columnwidth,keepaspectratio]{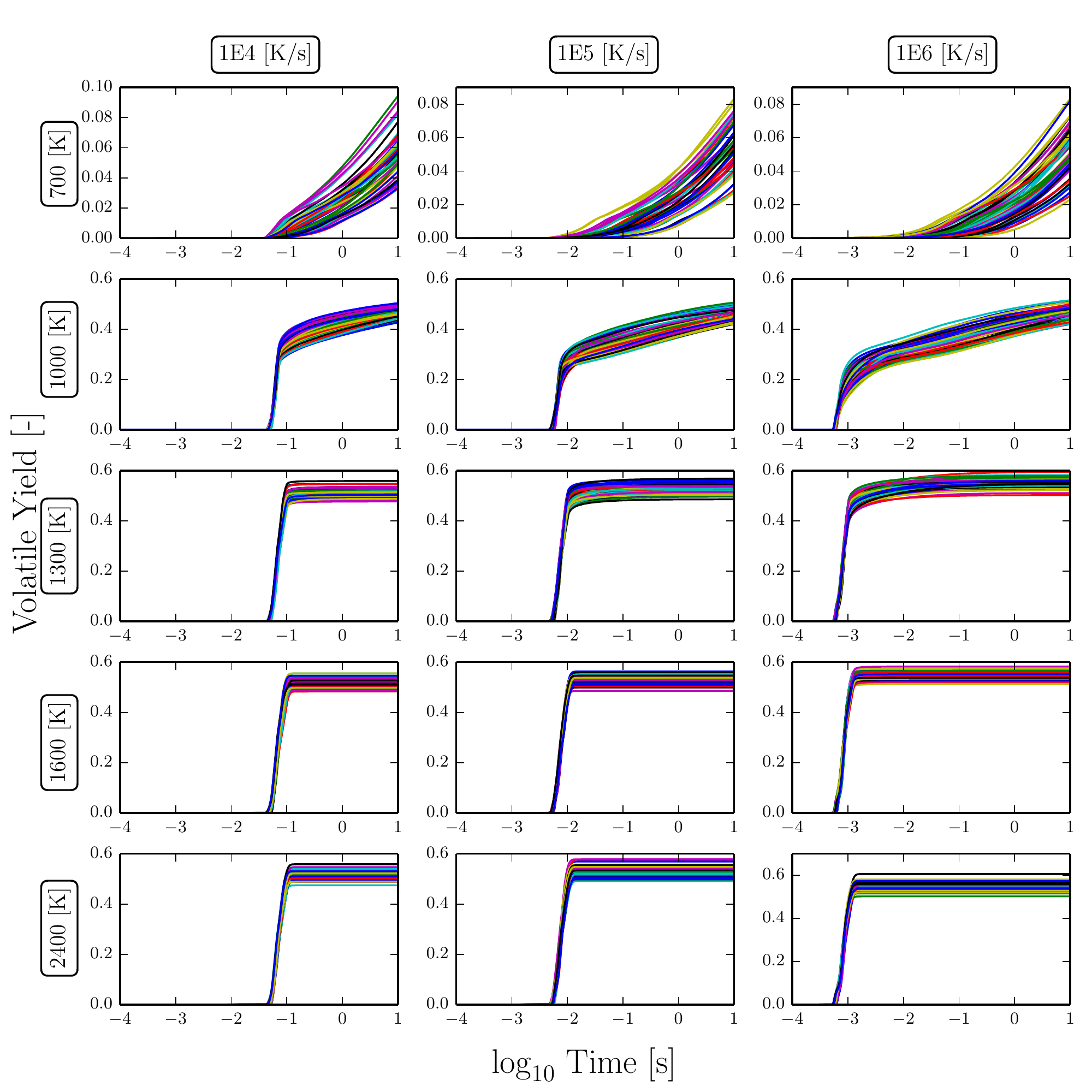}
\caption{For each of the 15 DOE conditions, Monte Carlo samples of the 13 uncertain model parameters input into the CPD calculations were used to create volatile yield traces demonstrating the effect of those uncertainties}
\label{fig:Uncertainty_Invest}
\end{figure}

The effect of uncertainty in the 13 CPD model parameters was characterized with an uncertainty analysis completed for the 15 DOE conditions by taking 1,000 random samples from the uncertain parameter-space, also known as a hypercube.
A sample of the volatile yield traces' uncertainty produced by this random sampling is shown in Figure~\ref{fig:Uncertainty_Invest}.

Clearly the parameter uncertainty causes significant effects within the CPD outputs.
To better quantify the uncertainty in the CPD results, the data for the ultimate volatile yield was tabulated into statistical information within Table~\ref{tab:uncertainstatstable}.
\begin{table}[t!]
\caption{Uncertainty statistics from 1,000 random samples of CPD's uncertain parameters arranged according to the DOE matrix in Table~\ref{tab:doetable} and characterized in Figure~\ref{fig:Uncertainty_Invest}, where uncertainty values refer to ultimate volatile yield}
\label{tab:uncertainstatstable}
\centering
\resizebox{0.9\columnwidth}{!}{
\begin{tabular}{ccc|ccc|ccc}
\toprule
\rowcolor{black!10} \multicolumn{9}{c}{\large \textbf{Full DOE}} \\
\rowcolor{black!20} \multicolumn{3}{c}{Mean} & \multicolumn{3}{c}{Std.} & \multicolumn{3}{c}{Std / Mean} \\
\midrule
0.0528 & 0.05348 & 0.05216 & 0.01540 & 0.01524 & 0.01576 & 0.2943 & 0.2945 & 0.3000 \\
0.4605 & 0.4650 & 0.4652 & 0.02300 & 0.02331 & 0.02261 & 0.04974 & 0.05044 & 0.04875 \\
0.5174 & 0.5325 & 0.5494 & 0.01908 & 0.01953 & 0.01881 & 0.03696 & 0.03670 & 0.03428 \\
0.5174 & 0.5331 & 0.5504 & 0.01861 & 0.02010 & 0.01956 & 0.03593 & 0.03777 & 0.03557 \\
0.5183 & 0.5329 & 0.5499 & 0.01908 & 0.01931 & 0.01888 & 0.03682 & 0.03623 & 0.03443 \\
\midrule
\rowcolor{black!20} \multicolumn{3}{c}{Min.} & \multicolumn{3}{c}{Max.} & \multicolumn{3}{c}{(Max. - Min.)/Mean} \\
0.02032 & 0.01922 & 0.01803 & 0.09839 & 0.09694 & 0.09776 & 1.493 & 1.5022 & 1.518  \\
0.3991 & 0.4050 & 0.4058 & 0.5270 & 0.5328 & 0.5345 & 0.2771 & 0.2765 & 0.2776 \\
0.4603 & 0.4754 & 0.4998 & 0.5696 & 0.5890 & 0.6078 & 0.2117 & 0.2135 & 0.1968 \\
0.4686 & 0.4764 & 0.5023 & 0.5751 & 0.5852 & 0.6057 & 0.2057 & 0.2044 & 0.1880\\
0.4603 & 0.4843 & 0.4880 & 0.5696 & 0.5900 & 0.6068 & 0.2110 & 0.1982 & 0.2165 \\
\midrule
\rowcolor{black!10} \multicolumn{9}{c}{\large Marginalize Holding-Temperature} \\
\rowcolor{black!20} \multicolumn{9}{c}{[ Max(Std/Mean) - Min(Std/Mean) ] / Mean(Std/Mean)} \\
\multicolumn{9}{c}{2.847 \quad 2.834 \quad 2.933} \\
\rowcolor{black!20} \multicolumn{9}{c}{[ Max((Max-Min)/Mean) - Min((Max-Min)/Mean) ] / Mean((Max-Min)/Mean)} \\
\multicolumn{9}{c}{2.683 \quad 2.722 \quad 2.774} \\
\midrule
\rowcolor{black!10} \multicolumn{9}{c}{ \large Marginalize Heating Rate} \\
\rowcolor{black!20} \multicolumn{9}{c}{[ Max(Std/Mean) - Min(Std/Mean) ] / Mean(Std/Mean) } \\
\multicolumn{9}{c}{0.01932 \quad 0.03416 \quad 0.07459 \quad 0.1055 \quad 0.06037}\\
\rowcolor{black!20} \multicolumn{9}{c}{[ Max((Max-Min)/Mean) - Min((Max-Min)/Mean) ] / Mean((Max-Min)/Mean)} \\
\multicolumn{9}{c}{0.01678 \quad 0.003766 \quad 0.08035 \quad 0.08882 \quad 0.08746} \\
\bottomrule
\end{tabular}}
\end{table}
The normalized standard deviation and normalized complete range provide quantitative measures of the direct effect of the model parameter uncertainties.
Comparing those same statistics after marginalizing over the hold-temperature or heating-rate shows that the effect of hold-temperature is at least an order of magnitude more significant than that of the heating-rate for the thermodynamic trends.
It should also be noted that the uncertainty in CPD parameters causes uncertainty in the volatile yields on the same order of magnitude as observed when varying the heating-rate.

\subsection{Sensitivity Analysis}
\label{subsec:sensitivity_analysis}

A baseline sensitivity analysis for the 13 model parameters in CPD was conducted exploring local and global sensitivities with a screening approach.
Local sensitivities were estimated by perturbing uncertain parameters by 0.5\% of their nominal value.
This was done for one parameter at a time and the effects of adding and subtracting this perturbation were averaged, a one-at-a-time sensitivity measure variation \cite{Hamby1994}.
Likewise, global sensitivity was assessed by moving one uncertain parameter at a time to the edge of its prior bounds (equivalent to taking an order of magnitude larger perturbation than the local sensitivity analysis).
The local and global sensitivity of each uncertain parameter for all 15 DOE conditions was calculated in order to account for the hold-temperature and heating-rate effects upon the sensitivities.
The absolute amount that the ultimate volatile yield changes due to the altered parameter value is considered the sensitivity for this study.
Sensitivities were normalized for each DOE condition so that the largest is scaled to unity.
This scaling allows simple identification of parameters with comparatively minor impacts upon CPD outputs.

After analyzing the local and global sensitivities, 5 of the 13 model parameters were deemed to contribute relatively minor amounts of uncertainty to the CPD calculations.
Relatively minor is quantified as at least two orders of magnitude less sensitive for all DOE conditions than each set of conditions' most sensitive parameter for both the local and global sensitivity.
These five parameters were A$_b$, $\sigma_b$, A$_g$, $\sigma_g$, and A$_{cr}$.  Thus, the majority of uncertainty in CPD calculations can be attributed to 8 instead of 13 parameters.
This allows less points to be tested when characterizing CPD's uncertainty, which ultimately will be used to assess the SBM's performance.
All five of the parameters that CPD was found to be less sensitive to are general CPD model parameters, not coal specific, which makes this sensitivity finding applicable to studies for other coal types.
It was noted that the local scaling of parameter $ac$ was larger than the global scaling due to prior uncertainty bounds assigned, yet CPD was still found to be relatively sensitive to this parameter's uncertainty.

\subsection{Scale-Bridging Decisions}
\label{subsec:SBdecisions}

From the analysis of the detailed physics model, CPD, knowledge has been gathered that can be used during the design of the SBM.
For the application-scale, capturing thermodynamic trends must be prioritized over kinetics.
This ordering was chosen because thermodynamics act on larger temporal-scales and accurately modeling the mass of gaseous fuel within the boiler is judged to be more significant than the rate of volatile production.
Following this line of thought, it will be a better approximation to neglect the heat-rate effect than the hold-temperature when creating the SBM.
This approximation is also justified by the fact that the hold-temperature was shown to be more influential on thermodynamic traits of CPD when comparing Figure~\ref{fig:ConstRUltYield} and \ref{fig:ConstTUltYield}.
Transporting the heating-rate history of each coal particle would mean additional modifications within the CFD code, so the ability to make this assumption is practical.

The other significant conclusion reached is that the primarily thermodynamic characteristic is only sensitive to 8 of the 13 uncertain CPD parameters, so less sampling will be necessary to sufficiently sample the parameter-space, or the 1,000 samples already collected are sufficient.
The uncertainty space exploration will be useful in calibration of the SBM throughout the upcoming model development.

\section{Scale-Bridging Model-Form}
\label{sec:SBM}

\begin{figure}[t!]
\centering
\includegraphics[width=0.6\columnwidth,keepaspectratio]{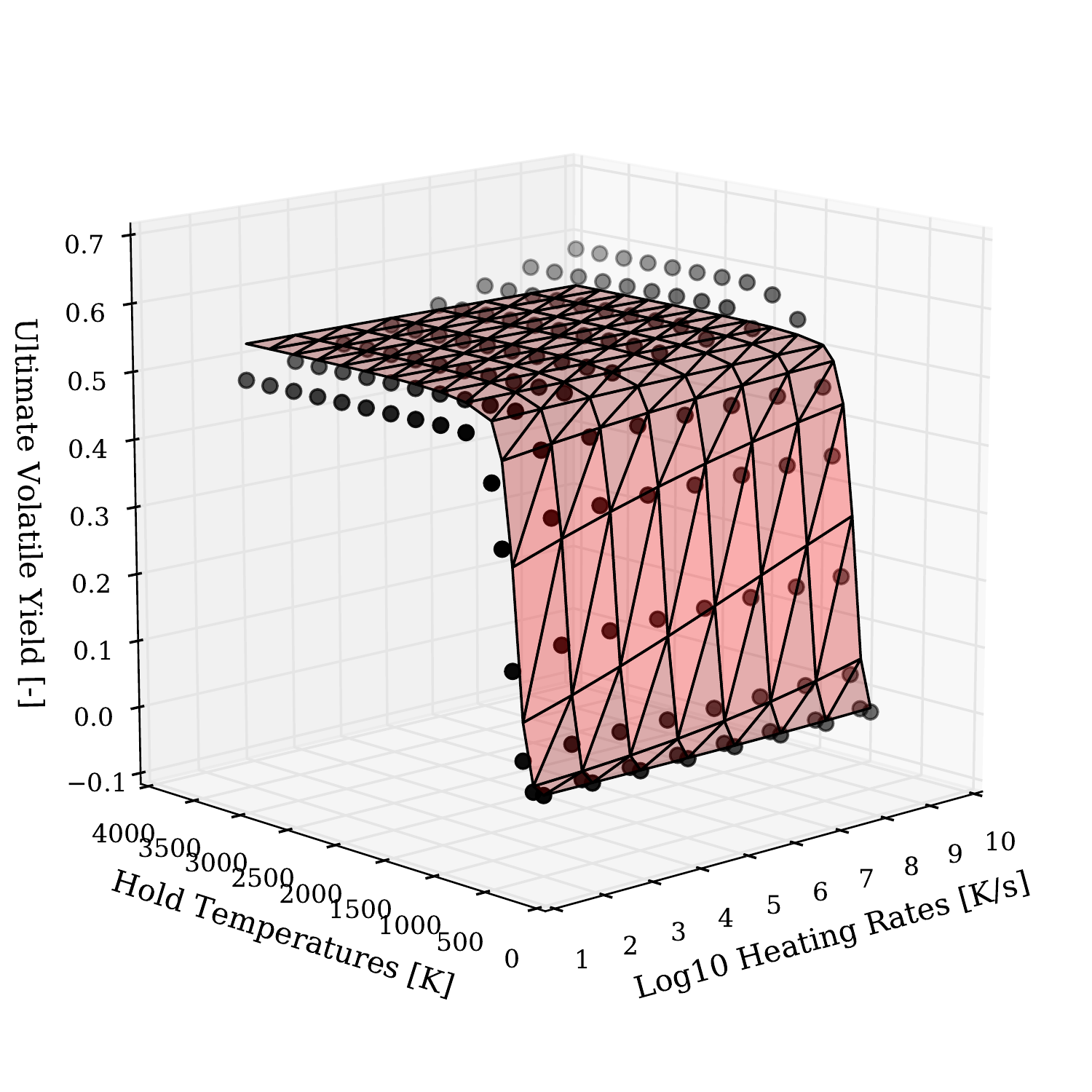}
\caption{Fitting a yield model Eq.~\eqref{eq:yield_model_eq} to CPD results across heating-rates and hold-temperatures.
Dots signify CPD data-points and the surface is an interpolation of the fitted yield model.}
\label{fig:yield_model_3d}
\end{figure}

\begin{figure}[t!]
\centering
\includegraphics[width=0.6\columnwidth,keepaspectratio]{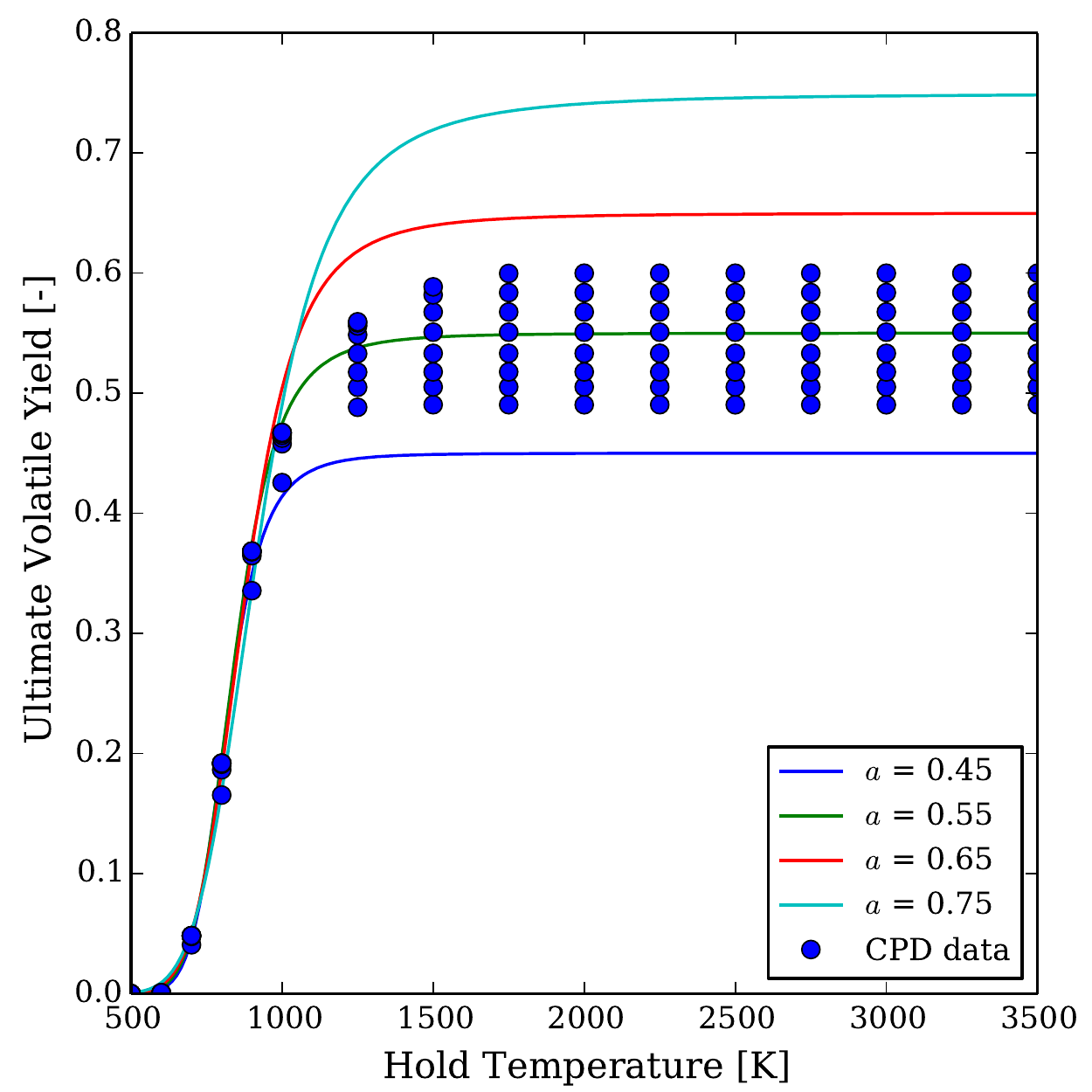}
\caption{Demonstrating effect of varying the high-temperature ultimate volatile yield parameter $a$ on the yield model Eq.~\eqref{eq:yield_model_eq}.
Each line signifies a different $a$ value and the dots are CPD data, also shown in Figure~\ref{fig:yield_model_3d}, where heating-rates and hold-temperatures were varied.
CPD data contains vertical spread due to varied heating rate from 1E2 - 1E9 K/s.}
\label{fig:yield_model}
\end{figure}

For the application conditions and scale of interest, the ultimate volatile yield has been determined to be major physics characteristic of interest.
When considering how to capture this effect, the approach of using a single first order reaction (SFOR) model, as presented by Biagini and Tognotti (2014) \cite{Biagini2014}, was considered.
\begin{align}\label{eq:SFOR}
\frac{\mathrm{d}V}{\mathrm{d}t} = A \exp(-E/T_P) \big(V_f - V)
\end{align}
Here $A \; [s^{-1}]$ is a pre-exponential factor, $E \; [K]$ is the activation temperature, $T_P \; [K]$ is the particle temperature, $V_f \; [-]$ is the ultimate volatile yield, and $V$ is the volatile yield.
Activation temperatures are equivalent to activation energies divided by the ideal gas constant.
What differentiates this SFOR model-form from previous uses of single-reaction devolatilization models, such as that by Badzioch and Hawksley (1970) \cite{Badzioch1970}, is that the ultimate volatile yield  is a function instead of a fixed value.
Biagini and Tognotti proposed an exponential form with the ultimate volatile yield having temperature dependence
\begin{align}
V_f = 1 - \exp\big(- DI \cdot \frac{T_P}{T_{st}}\big),
\end{align}
where $DI$ is the composition specific, dimensionless devolatilization index and $T_{st}$ is specified as the ``standard temperature" 1,223 K.
This specific model-form did not closely resemble the thermodynamic yield curve produced by CPD, as shown in Figure~\ref{fig:ConstRUltYield}, but further study of other forms appeared promising.
Alternative functional-forms can be explored through comparison with CPD results.  Fitting
\begin{align}\label{eq:yield_model_eq}
V_f = \frac{a}{2} \cdot \bigg(1 &- \tanh\big((b + c \cdot a) \, \cdot
( 590 - T_P )/T_P + (d + e \cdot a) \big)\bigg)
\end{align}
to CPD data can be seen in Figure~\ref{fig:yield_model_3d}-\ref{fig:yield_model} where $b = 14.26$, $c = -10.57$, $d = 3.193$, and $e = -1.230$ for Utah Sufco bituminous coal.
These values were found using a simplex minimization to fit the four curves shown in Figure~\ref{fig:yield_model} to the data extracted from CPD as shown across both explored dimensions in Figure~\ref{fig:yield_model_3d}.
Because the goal of this fitting was to match the CPD volatile yield data at low hold-temperatures, while allowing non-matching, higher ultimate volatile yields at high-temperatures, only data-points with hold-temperatures less than or equal to 1,000 K were used in the minimization.
The value 590 K within Eq.~\eqref{eq:yield_model_eq} was physically meant to describe the temperature at which devolatilization begins.
Although this volatilization temperature has currently been specified, it may be found to be coal dependent in the future.

From the three-dimensional view in Figure~\ref{fig:yield_model_3d}, it is evident that the yield model contained error due to not capturing the effect of the heating-rate, especially at higher temperatures.
Because the fitting included data from a range of heating-rates, the yield model was effectively fit to the center of the heating-rate range (1E5 - 1E6 K/s) and thus had the least error near those heating-rates.
Overall, it can be noted that this yield model satisfactorily captures CPD's yield trend across the necessary range of hold-temperatures.

Figure~\ref{fig:yield_model} demonstrates an attribute of Eq.~\eqref{eq:yield_model_eq} for which this model was specifically designed.
This model-form allows the ultimate volatile yield at high temperatures (relative to the application space of interest) be varied with parameter $a \; [-]$, while maintaining similar yields for lower hold-temperatures.
This characteristic is allowed because of the sparsity of devolatilization data for higher temperatures.
The optimal $a$ value found through fitting with CPD data less than or equal to 1,000 K was 0.54.
Ultimately, this uncertain variable will play a key role in exploring uncertainty in devolatilization for the application.
Within Figure~\ref{fig:yield_model} the uncertainty in the parameter $a$ can be noted to capture much of the same trend seen in the uncertainty within the CPD data due to the effect of the heating rate, with higher $a$ values appearing to correlate with faster heating-rates.
This variance also resembles the uncertainty in CPD outputs due to parameter uncertainty.

\begin{figure}[t!]
\centering
\includegraphics[width=0.4\columnwidth,keepaspectratio]{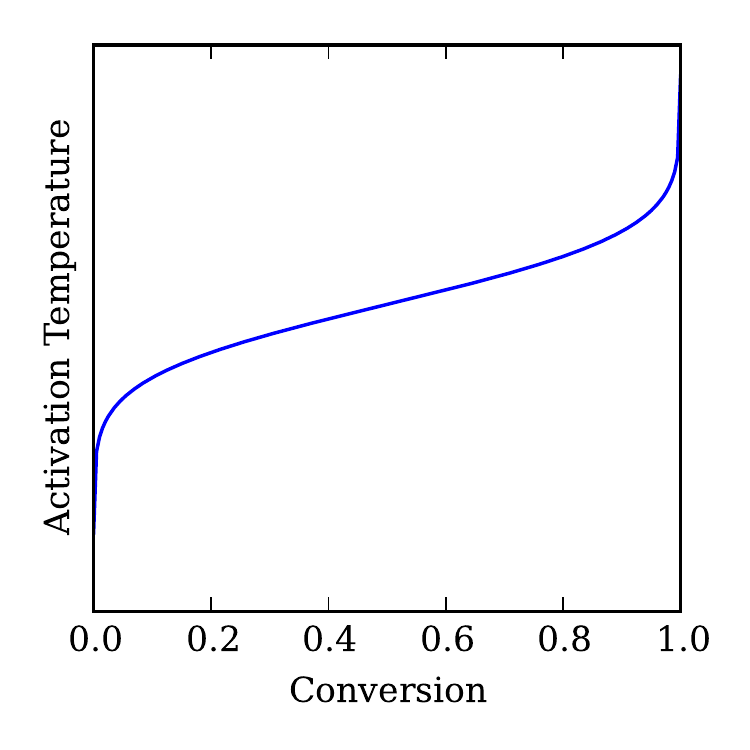}
\caption{Activation temperature's inverse cumulative distribution trace normalized to the amount of potential conversion}
\label{fig:energy_dist}
\end{figure}

The simple SFOR model-form Eq.~\eqref{eq:SFOR} was unable to satisfactorily capture the desired physical characteristics of CPD for the specified DOE conditions.
In order to better reproduce desired physical attributes, the concept of a distributed-activation energy model (DAEM) was incorporated into the reaction model.  DAEM is based upon the idea of representing devolatilization as an infinite series of parallel reactions \cite{Anthony1975}.
To model this concept it is assumed that there is a continuous distribution of activation temperatures and by evolving this distribution over time the effective activation temperature varies.
The integral form of Eq.~\eqref{eq:SFOR} with DAEM incorporated can be calculated as \cite{Smoot1985}
\begin{align}
\frac{V_f - V}{V_f} = \int_0^\infty \; \exp\bigg[-\int_0^t \; k \: \mathrm{d}t \bigg] f(E) \: \mathrm{d}E ,
\end{align}
where $k = A \exp(-E/T_P)$.
A Gaussian distribution was assumed to describe the distribution of the activation energies or
\begin{align}
f(E) = [\sigma_a(2 \pi)^{1/2}]^{-1} \exp \bigg(\frac{-(E-E_0)^2}{2 \sigma_a^2}\bigg).
\end{align}
One method of efficiently evaluating the DAEM model is to use the quadrature approximation to describe the distribution in terms of weights and abscissae \cite{Donskoi2000}.
Individual abscissae are evolved separately and then the final volatile yield is found by reapplying the weights and summing the then weighted abscissae.
Alternatively, the DAEM can be approximated by the activation temperature's inverse cumulative distribution function normalized by the full potential devolatilization conversion or
\begin{align}
Z = \text{max}\big(-4.0, \; \text{min}\big(&\sqrt{2.0} \cdot \erfinv (1.0 -
2.0\cdot(V_f - V)/a), \; 4.0\big)\big),
\end{align}
as illustrated in Figure~\ref{fig:energy_dist} \cite{Fletcher1992}.
Note that the yield model's high-temperature ultimate volatile yield $a$ was used as the measure of conversion extent for this implementation.
This distribution has been truncated to avoid numerical issues with the tails of the distribution.
With this method the effective activation temperature is initially low and as conversion progresses the activation temperature increases.
For devolatilization this causes reactions to initially accelerate quickly, but then decelerate while progressing towards completion.
The inverse cumulative distribution representation of the DAEM was selected for implementation due to its simple computation.

Mathematically the ultimate form of the SBM was enacted as
\begin{align}\label{eq:SRWY}
\frac{\mathrm{d}V}{\mathrm{d}t} &=
\begin{cases}
A \exp \bigg( \frac{-(E + \sigma_{a} Z)}{T_P} \bigg) (V_f - V), & \text{if } V_f - V \leq 0 \\
0, & \text{else}
\end{cases} ,
\end{align}
where the activation temperature $E \; [K]$, the distribution's standard deviation $\sigma_{a} \; [K]$, the pre-exponential parameter $A$, and the yield model's high-temperature ultimate volatile yield $a$ are free-parameters.
Equation~\eqref{eq:SRWY} is referred to hereafter as the single reaction with yield (SRWY) model-form.
This equation is explicitly presented as a conditional not because this affects calculations presented within the current analysis, but for the ultimate CFD application.
Without this conditional statement, reversed devolatilization, or condensation, could occur for particles moving from hot to cooler regions of the boiler.
Within a boiler coal particles are transported by turbulent eddies and likely experience heating and cooling events throughout their time in residence.
It is generally believed that coal particles experiencing cooling within the boils will not gain mass due to devolatilization.
This is because it is assumed that devolatilization is an irreversible phenomena in this environment.
Thus, the model-form was build to reflect the assumption of irreversibility.
Widening the perspective, once a coal particle reaches devolatilization thermodynamic equilibrium for a given temperature, no additional material will be devolatilized unless the particle experienced a higher-temperature that corresponds with a greater equilibrium volatile yield.
Thus, a coal particle will not devolatilize material when it is heated if it has already devolatilized to equilibrium at a higher temperature previously.
Beyond building this physical notion into the model-form, a few additional constraints were used throughout the model-form selection process.

Additional constraints considered during the model-form selection that ultimately led for Eq.~\eqref{eq:SRWY} included low-temperature limits, regions of greater emphasis for matching CPD, as well as allowing room for improvement at higher temperatures.
While leaving latitude for defining what constitutes low temperatures for coal devolatilization, it was generally conceived that minimal devolatization should occur in such regions.
While kinetic rates might allow material to leave a coal particle at low temperatures, physically there should not be sufficient driving force available.
The perceived low temperature behavior should be captured within the model-form of Eq.~\eqref{eq:SRWY} through the temperature dependent thermodynamically based ultimately volatile yield term within the driving force.
Then, above such low temperatures it is believed that CPD should accurately capture devolatilization for what will be considered the middle range of temperatures relevant to the application boilers.
Again the ultimate volatile yield model allows close matching of CPD within this temperature region due to it being fit to CPD data.
Due to the current lack of knowledge about how the devolatilization process behaves at high temperatures, flexibility was desired for such conditions.
Many options for approaching this uncertain high-temperature behavior exist and the current model-form is one that is simple to manipulate.
The four free-parameters can now be explored with a consistency analysis.

\section{Consistency Evaluation}
\label{sec:consistency_eval}

A first step in locating a set of parameters for the SRWY model-form, which can be used for the application, was to characterize a consistent space within the parameters' prior hypercube.
The idea of a consistent space comes from the methodology described within Feeley et al. (2004) \cite{Feeley2004}, where it was used to calibrate parameter values for the methane combustion reaction-set, GRI-Mech 3.0.
The basic premise of a consistent space can be described by the following equation.
\begin{align}\label{eq:consistencyEq}
(1-\gamma)l_i \leq M_i(\textbf{x}) - d_i \leq (1-\gamma)u_i \quad \text{for} \; i = 1,...,N_{QoI}
\end{align}
Here the model outputs $M$ at specified parameter inputs $\textbf{x}$ are compared directly to experimental data $d$.
This comparison proves consistency for a particular quantity of interest (QoI) $i$ if it is within the upper $u$ and lower $l$ error-bounds of the experimental data.
Model consistency is then achieved if consistency for all QoIs is found and the parameter ranges of the consistent sets overlap.
This model consistency or inconsistency can be further characterized by the decimal fraction that the error-bounds could be shrunk while maintaining consistency or the amount they could be expanded to reach consistency with the term $\gamma$.
The parameter hypercube is specified as
\begin{align}
\beta_p  \geq x_p \geq \alpha_p \quad \text{for} \; p = 1,...,n,
\end{align}
where $\beta$ and $\alpha$ designate the prior bounds for each parameter $n$.
If model consistency is found, it will correspond to a subspace of the hypercube.
If no consistent region is located, then the hypercube's bounds as well as the experimental error-bounds could be reevaluated for possible expansion.
A convenient method of approaching inconsistent systems is to look at unary and binary consistency or sensitivity to individual data-points for outliers.

For the current analysis the experimental data is produced by the detailed physics model, CPD.
The uncertainty in the QoIs, as quantified by the uncertainty exploration (Sec.~\ref{subsec:UQ}), will act as the error-bounds.
One thousand random samples, such as those previously shown in Figure~\ref{fig:Uncertainty_Invest}, were collected for use in defining QoI bounds.
This quantity of sampling is deemed sufficient considering CPD was found to be sensitive to only 8 of the 13 uncertain parameters (Sec.~\ref{subsec:sensitivity_analysis}).

\begin{figure}[t!]
\centering
\includegraphics[width=0.4\columnwidth,keepaspectratio]{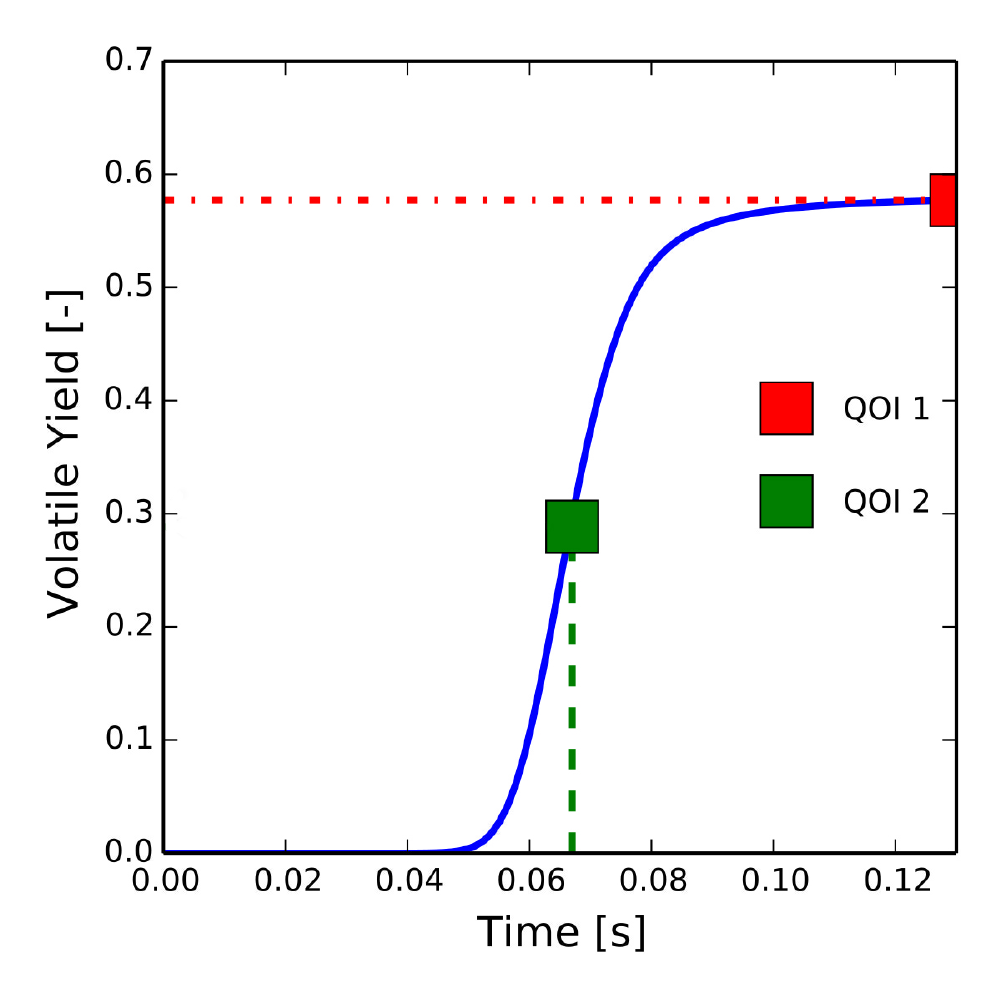}
\caption{Illustration of QoI definitions used within consistency analysis.
Blue line is a devolatilization volatile yield trace over time.
The red and green boxes demonstrates the position of first and second QoI which are the ultimate volatile yield and time to half the ultimate volatile yield, respectively.
Dashed and dashed-dotted lines illustrate where the boxes fall on the time scale and volatile yield.}
\label{fig:QoI_Visual}
\end{figure}

Defining QoIs is perhaps the most subjective component of the consistency analysis.
The simplest method of selecting a QoI is to use the ultimate quantity in which a prediction is desired.
An alternative selection procedure is to use features related to attributes deemed physical and/or necessary for accurate predictions of a physical phenomena, but which have no means of direct comparison.
The two QoI definitions chosen for determining the SRWY model-form's consistency are the ultimate volatile yield and the time to get to half the ultimate volatile yield.
The ultimate volatile yield is the quantity directly desired from the SRWY model-form and represents capturing thermodynamic trends of coal devolatilization relevant to the application.
Although kinetics were set to secondary importance when the heating-rate effects were deemed less significant in Sec.~\ref{subsec:SBdecisions}, roughly capturing devolatilization kinetics is still desired.
Thus, the second time-based QoI was selected to enable the SRWY model-form to roughly estimate CPD's kinetic behaviors.
The QoIs are visually depicted within Figure~\ref{fig:QoI_Visual}.

Using two QoIs for each of the 15 DOE conditions results in a total of 30 QoIs, as shown in Table~\ref{tab:QoItable}, that must be simultaneously satisfied for consistency.
\begin{table}[t!]
\caption{QoI ranges for 15 DOE conditions, where QoIs are the ultimate volatile yield and the time to get to half the ultimate volatile yield}
\label{tab:QoItable}
\centering
\resizebox{0.9\columnwidth}{!}{
\begin{tabular}{c|cc|cc|cc}
\toprule
\rowcolor{black!10} \multicolumn{7}{c}{\large \textbf{QoIs}} \\
\midrule
\rowcolor{black!20} & \multicolumn{2}{c}{DOE 1} & \multicolumn{2}{c}{DOE 2} & \multicolumn{2}{c}{DOE 3} \\
$\frac{1}{2}$ Ult. Vol. Yield Time [s] & \multicolumn{2}{c}{8.099e-01 -- 2.857e+00} &  \multicolumn{2}{c}{7.544e-01 -- 2.896e+00} & \multicolumn{2}{c}{7.627e-01 -- 2.897e+00} \\
Ult. Vol. Yield [-] & \multicolumn{2}{c}{2.032e-02 -- 9.839e-02} & \multicolumn{2}{c}{1.922e-02 -- 9.694e-02} & \multicolumn{2}{c}{1.803e-02 -- 9.776e-02} \\
\midrule
\rowcolor{black!20} & \multicolumn{2}{c}{DOE 4} & \multicolumn{2}{c}{DOE 5} & \multicolumn{2}{c}{DOE 6} \\
$\frac{1}{2}$ Ult. Vol. Yield Time [s] & \multicolumn{2}{c}{6.312e-02 -- 7.019e-02} &  \multicolumn{2}{c}{6.933e-03 -- 9.686e-03} & \multicolumn{2}{c}{9.058e-04 -- 3.728e-03} \\
Ult. Vol. Yield [-] & \multicolumn{2}{c}{3.991e-01 -- 5.270e-01} & \multicolumn{2}{c}{4.050e-01 -- 5.328e-01} & \multicolumn{2}{c}{4.058e-01 -- 5.345e-01} \\
\midrule
\rowcolor{black!20} & \multicolumn{2}{c}{DOE 7} & \multicolumn{2}{c}{DOE 8} & \multicolumn{2}{c}{DOE 9} \\
$\frac{1}{2}$ Ult. Vol. Yield Time [s] & \multicolumn{2}{c}{6.402e-02 -- 7.381e-02} &  \multicolumn{2}{c}{7.035e-03 -- 8.073e-03} & \multicolumn{2}{c}{7.867e-04 -- 1.155e-03} \\
Ult. Vol. Yield [-] & \multicolumn{2}{c}{4.603e-01 -- 6.500e-01} & \multicolumn{2}{c}{4.754e-01 -- 6.500e-01} & \multicolumn{2}{c}{4.998e-01 -- 6.500e-01} \\
\midrule
\rowcolor{black!20} & \multicolumn{2}{c}{DOE 10} & \multicolumn{2}{c}{DOE 11} & \multicolumn{2}{c}{DOE 12} \\
$\frac{1}{2}$ Ult. Vol. Yield Time [s] & \multicolumn{2}{c}{6.456e-02 -- 7.415e-02} &  \multicolumn{2}{c}{7.120e-03 -- 8.068e-03} & \multicolumn{2}{c}{7.881e-04 -- 1.153e-03} \\
Ult. Vol. Yield [-] & \multicolumn{2}{c}{4.500e-01 -- 7.000e-01} & \multicolumn{2}{c}{4.500e-01 -- 7.000e-01} & \multicolumn{2}{c}{4.500e-01 -- 7.000e-01} \\
\midrule
\rowcolor{black!20} & \multicolumn{2}{c}{DOE 13} & \multicolumn{2}{c}{DOE 14} & \multicolumn{2}{c}{DOE 15} \\
$\frac{1}{2}$ Ult. Vol. Yield Time [s] & \multicolumn{2}{c}{6.443e-02 -- 7.392e-02} &  \multicolumn{2}{c}{7.076e-03 -- 8.075e-03} & \multicolumn{2}{c}{7.848e-04 -- 1.152e-03} \\
Ult. Vol. Yield [-] & \multicolumn{2}{c}{4.500e-01 -- 7.600e-01} & \multicolumn{2}{c}{4.500e-01 -- 7.600e-01} & \multicolumn{2}{c}{4.500e-01 -- 7.600e-01} \\
\bottomrule
\end{tabular}}
\end{table}
Initially, all QoI values were strictly based upon uncertainty quantified in the uncertainty exploration of CPD, except for apparent outliers which were discarded.
This was altered to accommodate greater perceived uncertainty that CPD does not take into account.
There are high-temperature ultimate volatile yield data reported to be higher than what CPD predicts \cite{Neoh1984}.
An additional high-temperature data-point that could be considered is the sublimation point of graphite which appears to have many caveats but is roughly estimated to be approximately 3,950 K \cite{Abrahamson2003}.
This potential discrepancy with CPD is believed to be due to such data not being taken into account during its formulation.
In order to allow the SRWY model-form to reflect such high-temperature data, the ultimate volatile yield bounds for the DOE conditions with hold-temperatures 1,600 K and 2,400 K were enlarged to reflect the perceived potential span.
Through the course of exploratory consistency tests, it was deemed that the upper bounds of the ultimate volatile yield QoIs for DOE conditions at 1,300 K were limiting the higher temperature DOE conditions' ability to reach higher ultimate volatile yields with the current model-form.
Thus, these bounds were also raised for all three heating-rates.
The final alteration to the QoIs was to increase the temporal QoIs' upper bounds by multiplying them by 1.3 for the DOE conditions with 1E6 K/s heating-rate and hold-temperatures of 1,300 K, 1,600 K, and 2,400 K.
These bounds were extended to allow greater amounts of consistency across all DOE conditions.
With high temperatures and fast heating-rates, the kinetic timescales of those three DOE conditions should have minimal affect upon the application simulations.

Once QoI definitions were established and values set, random samples of the SBM's free parameters were tested for simultaneous consistency across all QoIs and $\gamma$ values collected for consistent samples.
Where typically $\gamma$ values indicate the ability to shrink data error-bounds, for this model-form uncertainty application a more useful interpretation of $\gamma$ is that it is indicative of how true the SBM fits the center of the QoIs' uncertainties.
In order to capture the full spectrum of CPD uncertainty, a range of $\gamma$ values would be required.

Even with these relatively simple QoIs there is ambiguity about their definition and how they are enacted.
For each SRWY model-form trace the ultimate volatile yield and time it took the SRWY model-form trace to get to half that ultimate volatile yield are compared with the CPD ranges of uncertainty for those same quantities.
An alternative comparison that could be explored would be to check if the SRWY model-form's trace passed through the volatile yield range equal to half the uncertain CPD ultimate volatile yield range within CPD's time frame for reaching half its ultimate volatile yield.
This could be visually conceived as checking that a SWRY trace line passes through a CPD defined QoI rectangle.

A few numerical methods are necessary for executing this analysis.
SRWY model-form temporal traces are evolved in time using an adaptive Huen-Euler integration scheme.
The adaptive method was found to handle the system's stiffness while remaining computationally efficient.
On a logarithmic time-scale between 1E-5 and 10 seconds, 200 points are saved for each volatile yield trace.
Values representing the time to half the ultimate volatile yield are then located using a second-order Newton polynomial interpolation.
This polynomial was chosen due to the curved nature of the volatile yield traces in the desired temporal periods.
For random number generation the random.random function from the Python Numpy library was used to generate random samples from a continuous uniform distribution \cite{Scipy}.

\section{Consistency Analysis}
\label{sec:consistency_analysis}

\begin{figure}[t!]
\centering
\includegraphics[width=0.6\columnwidth,keepaspectratio]{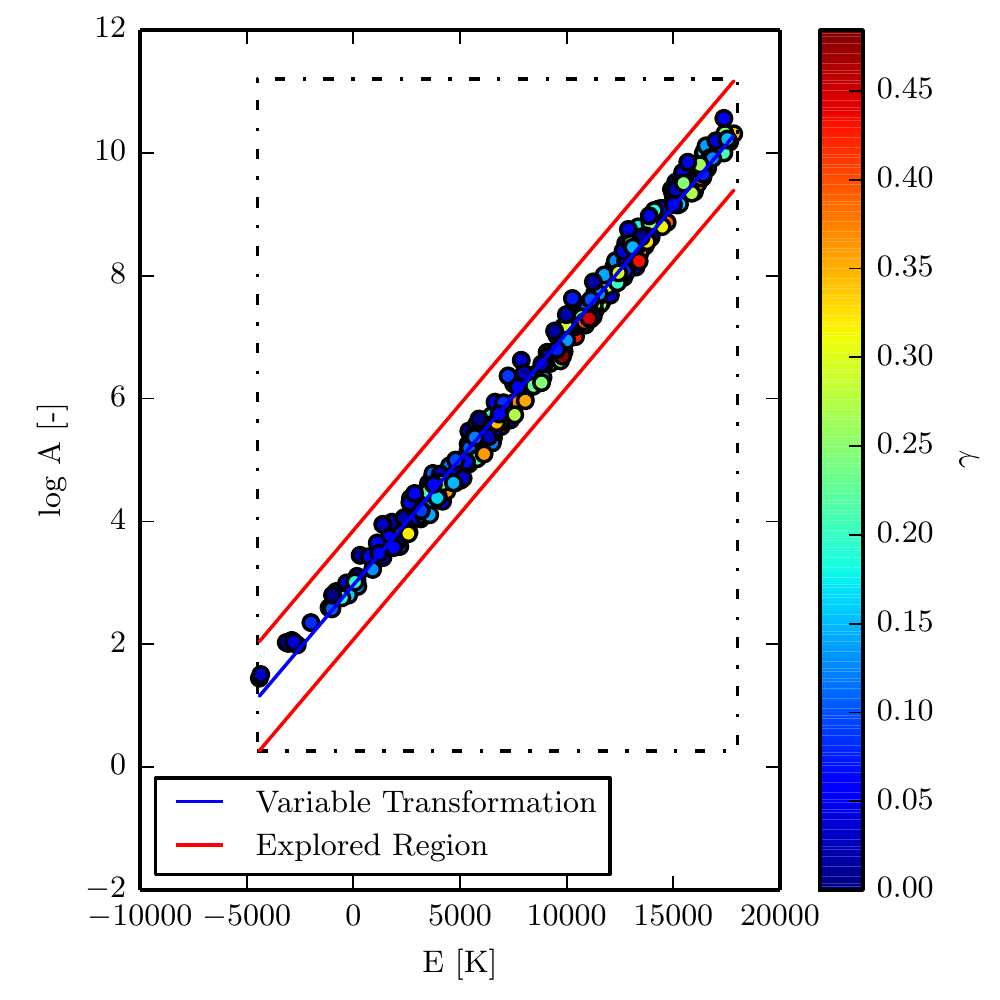}
\caption{Visualization of the variable transformation used for efficient exploration of the parameter-space.
Dots represent consistent points in the parameter-space and their color designate respective $\gamma$ values.
The variable transformation is shown through the blue line and the red lines indicate the region explored.}
\label{fig:var_trans_fig}
\end{figure}

\begin{figure}[t!]
\centering
\begin{subfigure}{0.7\columnwidth}
\includegraphics[width=\columnwidth,keepaspectratio]{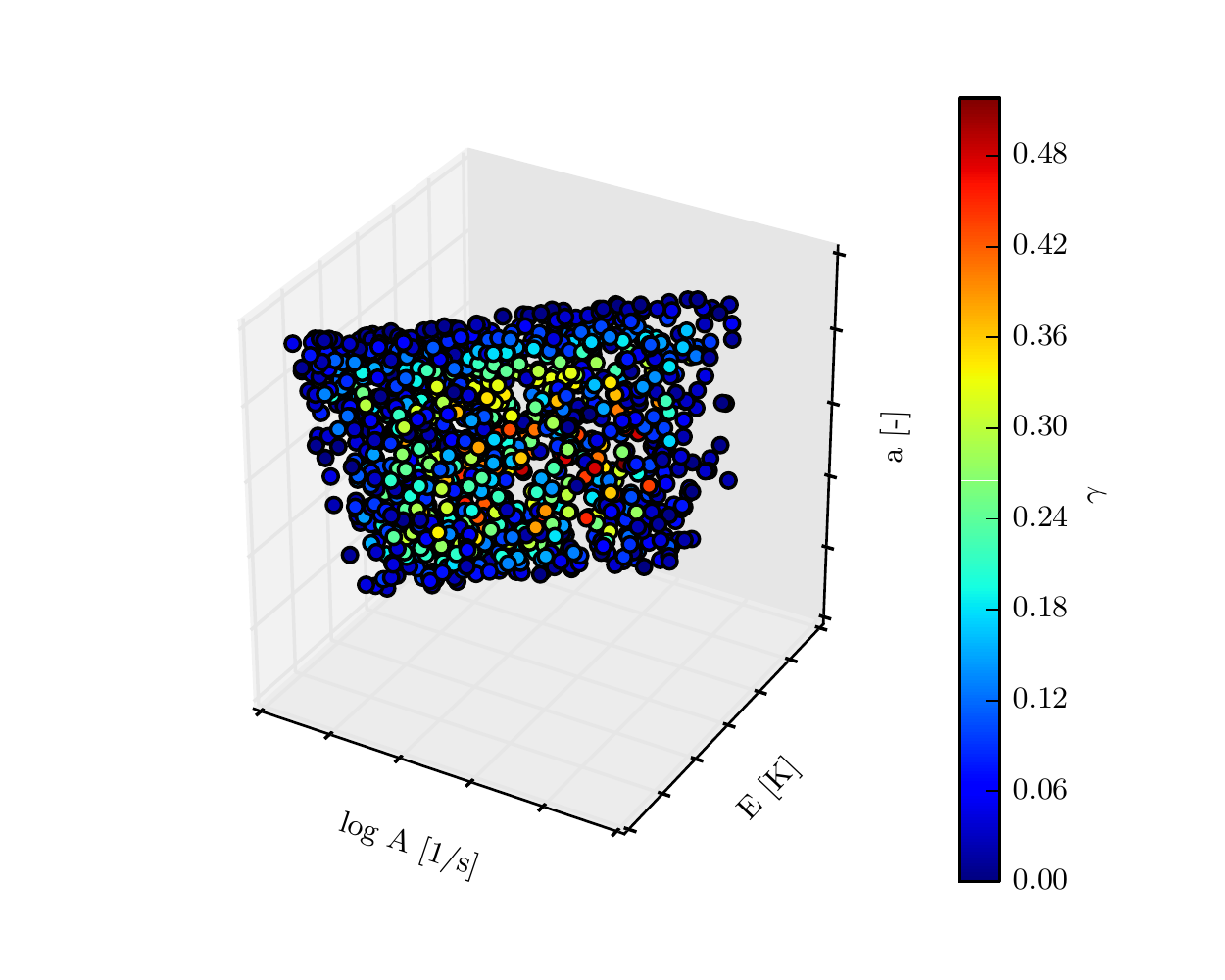}
\caption{}
\end{subfigure}
\begin{subfigure}{0.5\columnwidth}
\includegraphics[width=\columnwidth,keepaspectratio]{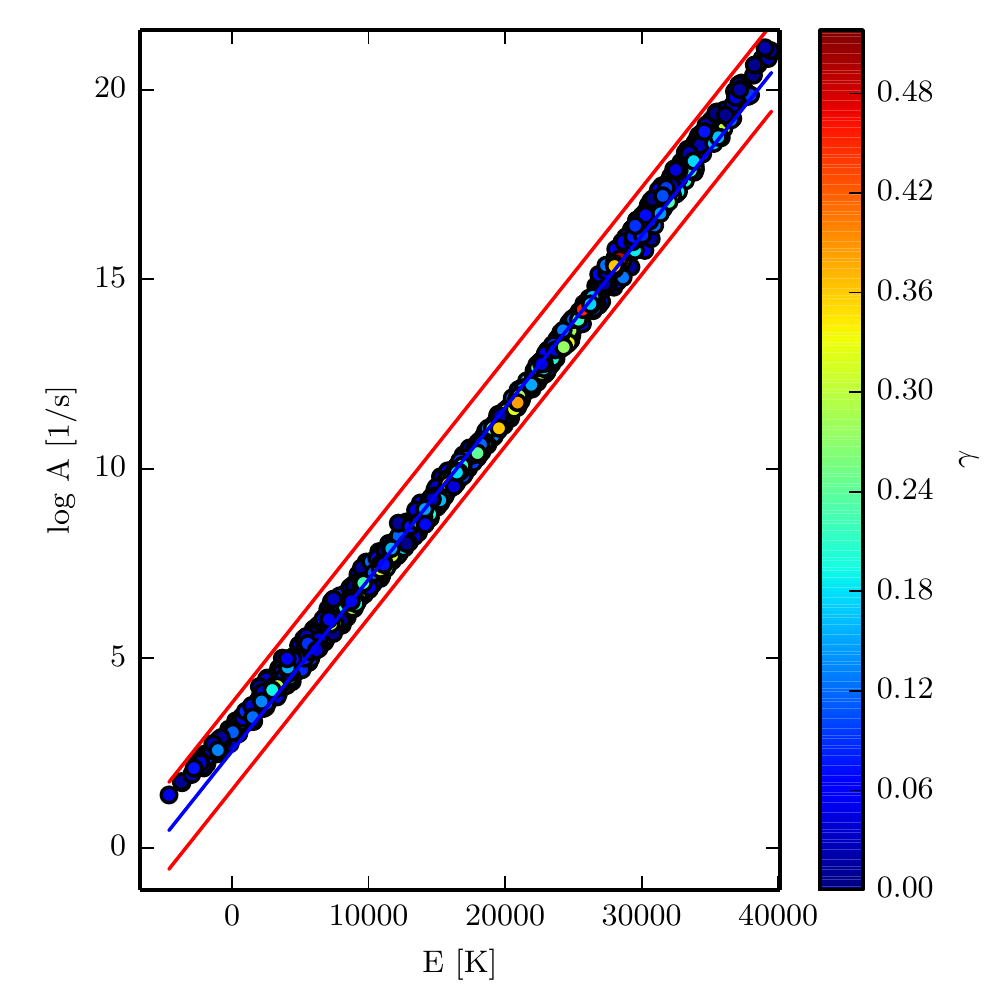}
\caption{}
\end{subfigure}
\caption{Marginal views of the SFWY model-form's four-dimensional parameter-space.
Dots signifying parameter-sets consistent with CPD uncertainty are shown in plot (a) across three parameter dimensions: $\log A$, $E$, and $a$.
Plot (b) shows a two-dimensional view of the consistent points across $\log A$ and $E$ space.
The dot color corresponds to the respective $\gamma$ value.}
\label{fig:3dConsistent}
\end{figure}

From visualizing a sample of consistent points in Figure~\ref{fig:var_trans_fig}, it was evident that exploring a transformed parameter-space would increase search efficiency.
A simple linear transformation based on the apparent linear correlation between $E$ and $\log_{10}(A)$ values was used:  $\log_{10}(A) = \text{slope} \cdot E \, + \, \text{intercept}$.
Instead of exploring the parameter-spaces of $A$ and $E$, the space of $E$ and the transformation's intercept can be explored once a slope was fitted to consistent points found during an initial search.
The greater efficiency of exploring the transformed space is demonstrated in Figure~\ref{fig:var_trans_fig}.
The dash-dotted rectangular region represents the pure parameter hypercube that would be explored if a variable transformation was not used and the region between the two red lines is the transformed space that can be explored more efficiently.
Throughout the process of refining the consistency search and regions of exploration, the transformation continuously evolved based upon the accumulation of additional data.
A more refined search of the transformed $E$ versus $\log_{10}(A)$ space is shown in Figure~\ref{fig:3dConsistent}.

\begin{figure}[t!]
\centering
\begin{subfigure}{0.5\columnwidth}
\includegraphics[width=\columnwidth,keepaspectratio]{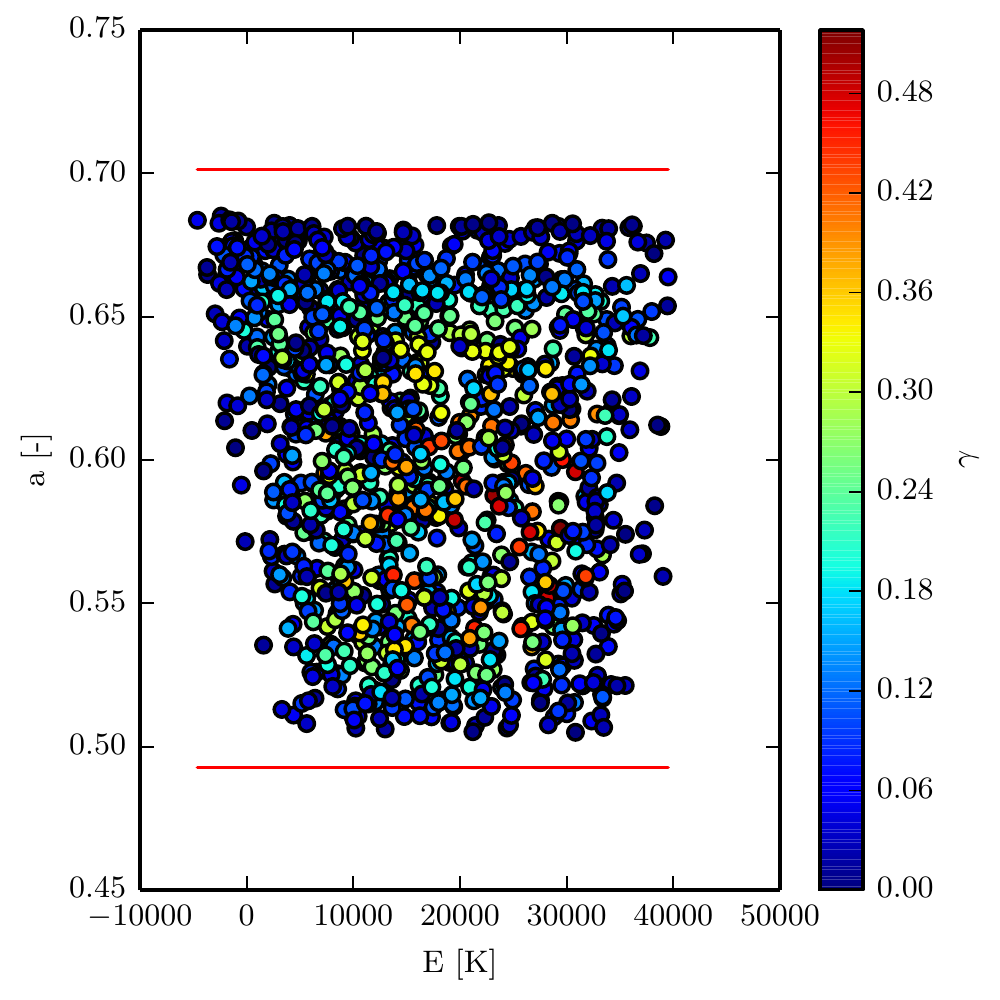}
\caption{}
\end{subfigure}
\begin{subfigure}{0.5\columnwidth}
\includegraphics[width=\columnwidth,keepaspectratio]{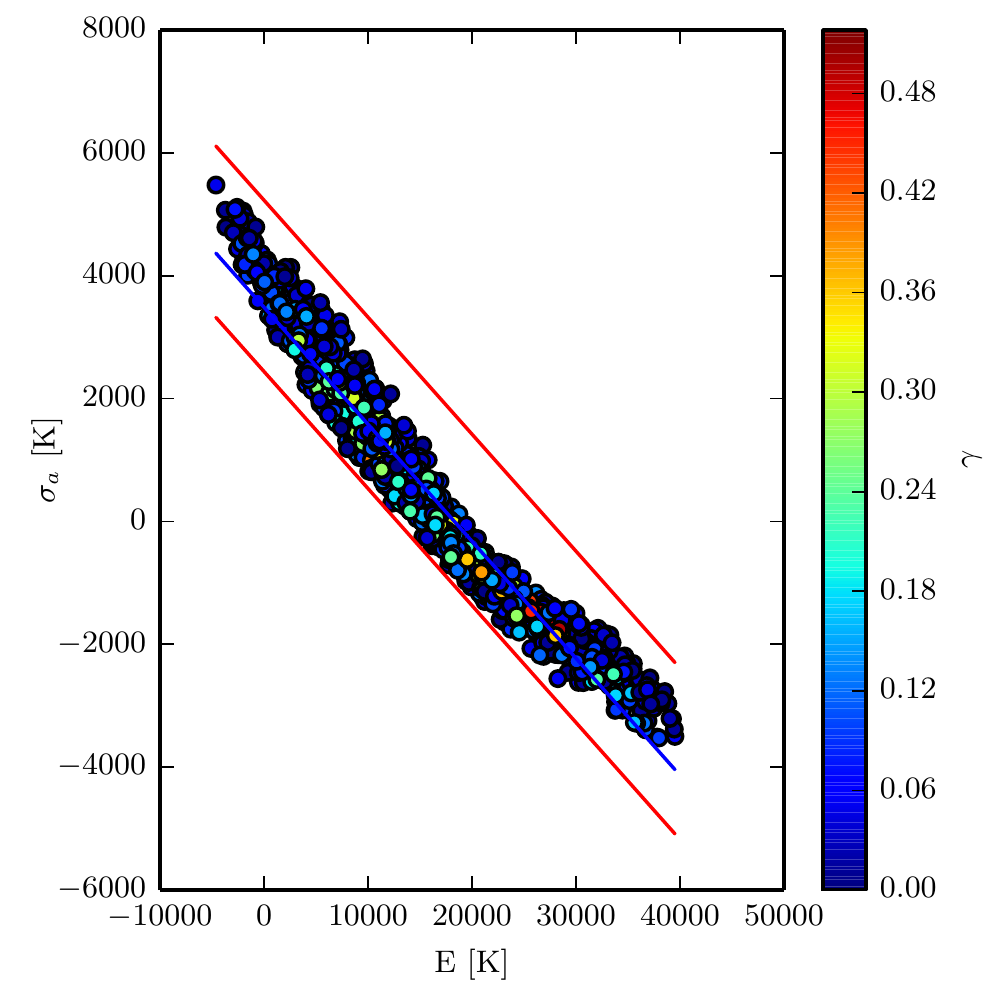}
\caption{}
\end{subfigure}
\caption{Visualization of two-dimensional parameter-spaces explored during the search of the four-dimensional parameter-space.
Plot (a) shows parameter $a$ versus parameter $E$ and plot (b) likewise shows the transformed space explored for parameter $\sigma_a$ versus $E$.
The blue line indicates the axis for transformed parameter-space and red lines indicate the explored region of parameter-space.}
\label{fig:other_trans}
\end{figure}

Linear transformed spaces for $a$ and $\sigma_a$, as shown in Figure~\ref{fig:other_trans}, were also used to minimize computational costs.
Although initially $a$ appeared to benefit from a variable transform, later this transformation was found to be unnecessary and the non-transformed parameter-space was used.
The consistent points shown in Figure~\ref{fig:3dConsistent}-\ref{fig:other_trans} were generated by taking 50,000 random samples from a four-dimensional transformed parameter-space.
Of those random samples, 1,244 consistent parameter-sets were found or approximately 2.5\%.

By considering Figure~\ref{fig:3dConsistent}-\ref{fig:other_trans} the consistent space can be noted to have many interesting characteristics.
The spread of the consistent space is relatively narrow across many two-dimensional visualizations or marginals of the data, demonstrating that the utility of variable transformations may encompass more than just an increase in search efficiency.
Bounds appear to exist for many of the parameters and shift throughout the four-dimensional space.
For instance parameter $a$ seems limited to a range between 0.5 and 0.69, but the lower bound shifts upwards for the extreme $E$ values.

Parameters $E$, $A$, and $\sigma_a$ appear to be highly correlated, as could be deduced by the variable transformations used to explore their spaces.
Such correlation was expected with fast heating rates because the coal was effectively experiencing a fixed temperature for a large temporal portion of the devolatilization.
Looking back at Eq.~\eqref{eq:SRWY} it can be shown that for a constant temperature the equation has non-unique solutions or correlation between three parameters.
\begin{align}
K = \text{Constant} &= A_1 \exp\big(-(E_{1} + \sigma_{a,1}Z)/T_P\big) \notag \\
\phantom{K = \text{Constant}} &= A_2 \exp\big(-(E_{2} + \sigma_{a,2}Z)/T_P\big) \notag \\
\ln(A_1) \, - \, \ln(A_2) &= \frac{E_{1} + \sigma_{a,1}Z}{T_P} - \frac{E_{2}  + \sigma_{a,2}Z}{T_P}
\end{align}
With this strong correlation the three parameters' bounds are also interdependent.
The  activation temperature $E$ covered the entire parameter region explored, but tapered off in the number of consistent points found and respective $\gamma$ values of those points in the limits of its explore region -9,000 K to 40,000 K.
Because of the non-unique behavior, it is not possible to deduce which of the three correlated parameters limited the span of the consistent space.

\begin{figure}[t!]
\centering
\includegraphics[width=0.5\columnwidth,keepaspectratio]{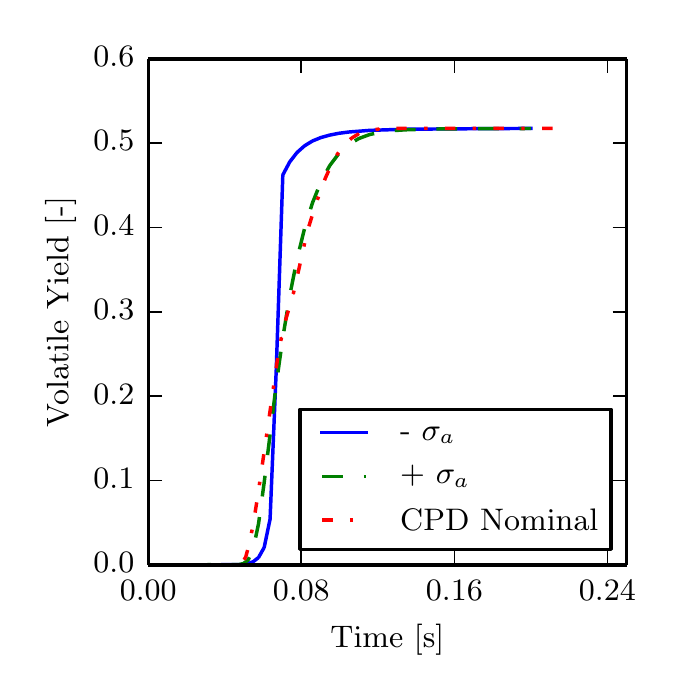}
\caption{Comparing volatile yield traces of CPD and the SRWY model-form with a hold-temperature of 2,400 K and heating-rate of 1E4 K/s.
The CPD trace (red dash-dotted line) was calculated with nominal parameter values, while the SRWY model-form traces were calculated with consistent parameter-sets.
One SRWY model-form trace utilizing a consistent parameter set with the standard temperature distribution ($+\sigma_a$) is shown as the green dashed line and the other using an inverted temperature distribution ($-\sigma_a$) is shown as the blue line.}
\label{fig:invert_energy}
\end{figure}

Another interesting feature of the SRWY model-form found through the consistency search was the ability to remain consistent while inverting the temperature distribution.
In Figure~\ref{fig:other_trans} it can be noted that consistent $\sigma_a$ values become negative when the activation temperature $E$ surpasses approximately 18,000 K.
The negative sign should not be thought of as part of the $\sigma_a$ value because a negative standard deviation is not possible.
Instead, looking back at Eq.~\eqref{eq:SRWY} and Figure~\ref{fig:energy_dist}, a negative value is indicative of inverting the activation temperature distribution, causing high activation temperatures to initially control the reaction and low activation temperatures to be operating when high amounts of conversion have occurred.
This change in the temperature distribution should significantly alter the kinetic trends, yet evidently does so in a manner that maintains consistency.
A visual comparison of the effect of inverting the temperature distribution can be seen in Figure~\ref{fig:invert_energy}, where two traces generated with consistent SRWY parameter-sets, but with opposite activation temperature distribution orientations, are compared with an equivalent nominal CPD trace.

It is evident that the original orientation of the temperature distribution produces a trace that bears characteristics more similar to that of the CPD trace.
The trace from the inverted temperature distribution has a sharper slope, but does level-off at the desired volatile yield.
While consistent parameter-sets with inverted temperature distributions could be discounted from further consideration due to the poor shape characteristics, they will be retained presently due to fulfilling the current consistency criteria.
If additional or redefined QoIs were used in the future, points with negative activation temperature distributions could justly be removed.
This issue illustrates the distinction between adequacy and credibility.
While the current consistency QoIs define model adequacy, they do not ensure credible solutions.

\begin{figure}[t!]
\centering
\includegraphics[width=0.7\columnwidth,keepaspectratio]{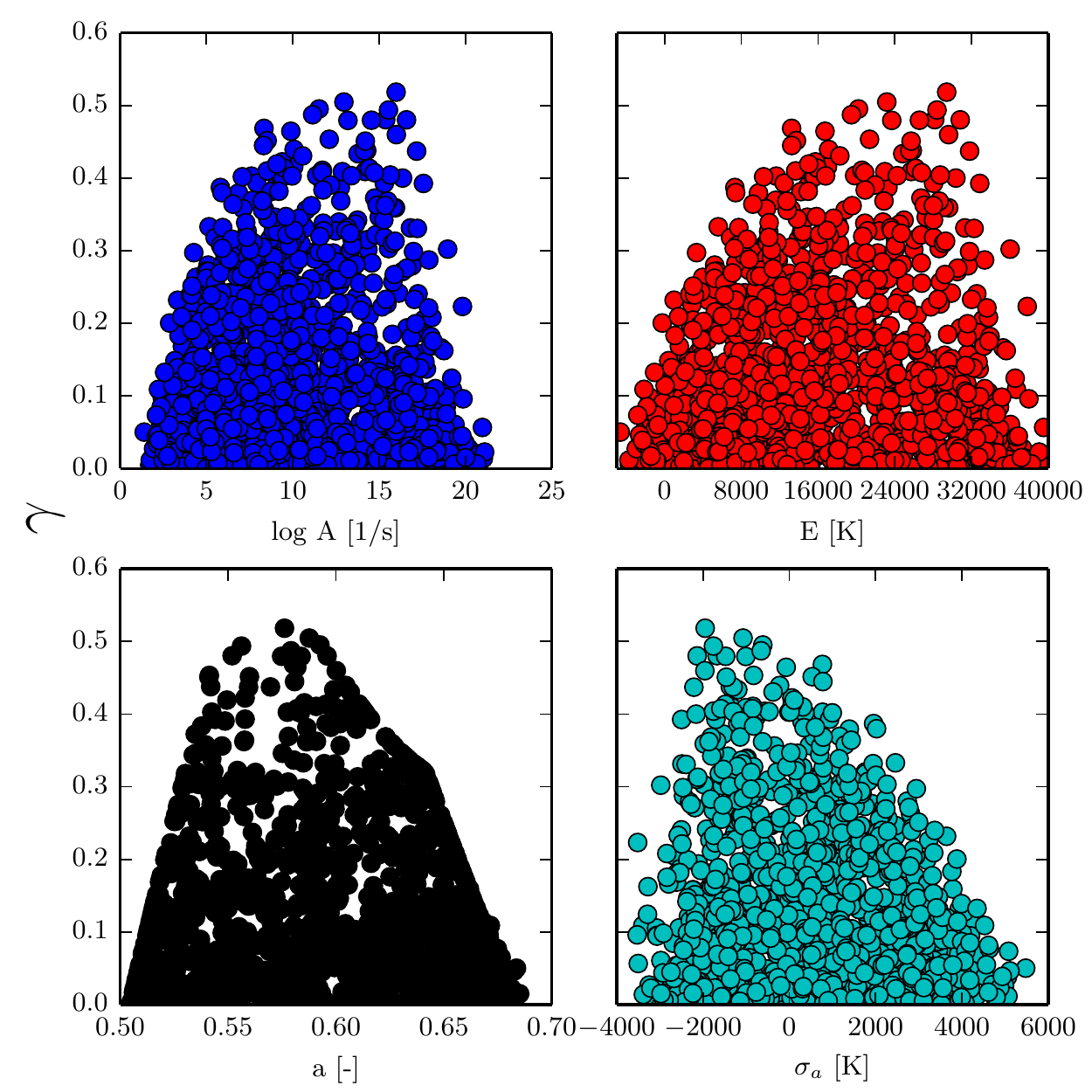}
\caption{Scatter plots showing how the $\gamma$ values of  consistent parameter-sets are distributed across the SRWY model-form's free-parameters.}
\label{fig:Gamma_Correlations}
\end{figure}

Throughout visualizations of consistent parameter-sets (Figure~\ref{fig:var_trans_fig}-\ref{fig:other_trans}), it is evident that $\gamma$ values associated with those consistent points are not continuously distributed across the four-dimensional space.
It appears that parameter-sets with higher $\gamma$ values typically lie within interior regions of the parameter-space and that the bounding regions of the consistent parameter-space have low gamma values as would be expected.
Exactly how the $\gamma$ values are distributed across each of the four uncertainty SRWY parameters can be visualized within Figure~\ref{fig:Gamma_Correlations}.

All four uncertain parameters appear to have reasonably well defined distributions indicating that the parameter-space exploration was sufficient.
Truncation of the edges of the distributions occurs, but should not significantly alter the distribution's appearance.
The high-temperature ultimate volatile yield $a$ correlates to $\gamma$ with a distribution that reaches a maximum between an $a$ value of 0.55 and 0.6.
This is not surprising considering CPD predicted a similar range, as was seen in Figure~\ref{fig:yield_model}.
This distribution also has clear bounds that appear almost smooth even with the limited sample points.
The most consistent parameter set or that with the highest $\gamma$ value found corresponded to $\log_{10}(A) = 15.980 \, s^{-1}$, $E = 29,400 \, K$, $\sigma_a = -1,950 \, K$, and $a = 0.576$.
This set does not represent the largest $\gamma$ possible for the SWRY model-form due to the use of random sampling, but acts as an estimate of the region of highest consistency.
If consistent points with inverted temperature distributions are discounted, the most consistent set would be $\log_{10}(A) = 8.331 \, s^{-1}$, $E = 13,240 \, K$, $\sigma_a = 767.6 \, K$, and $a = 0.580$.

\begin{figure}[t!]
\centering
\includegraphics[width=0.9\columnwidth,keepaspectratio]{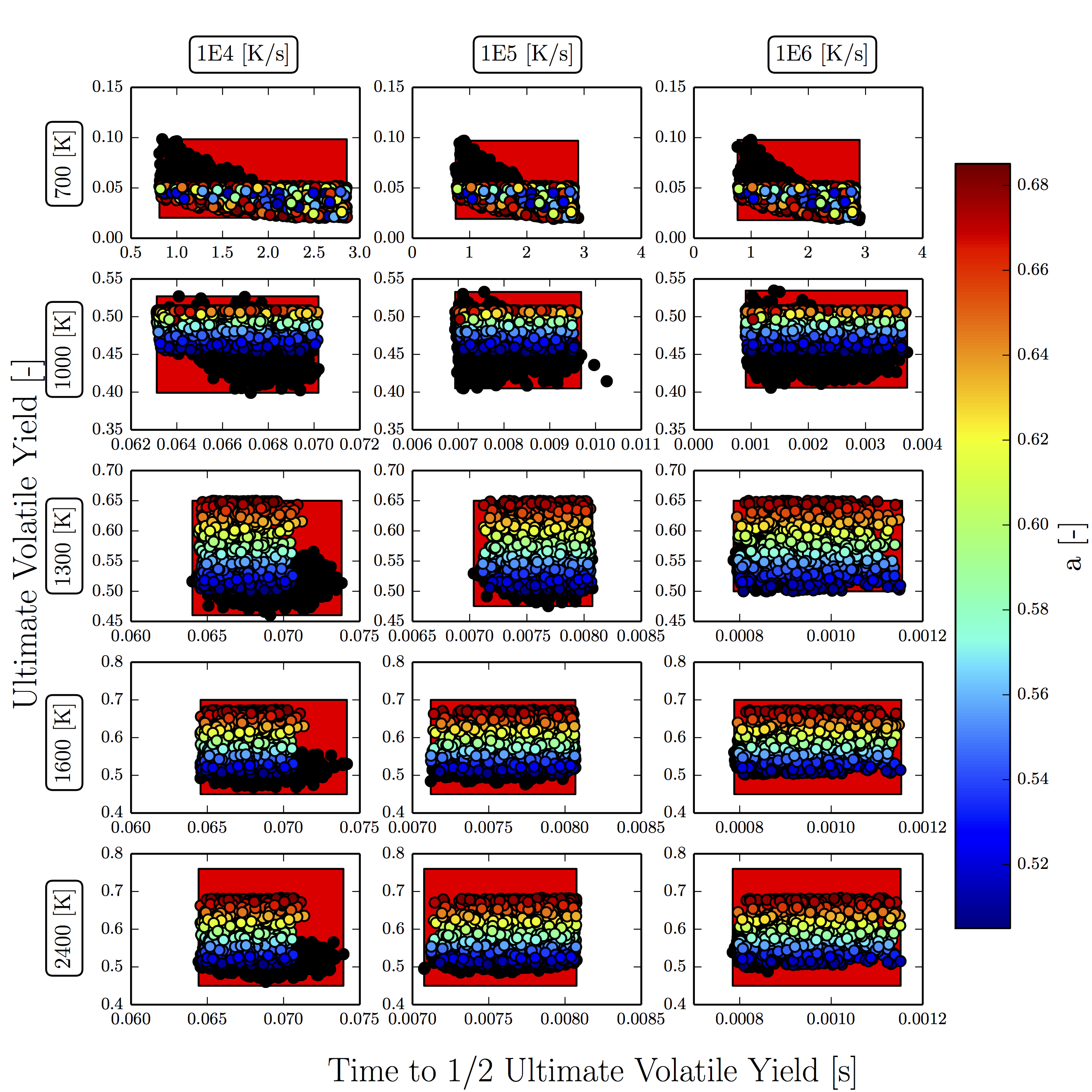}
\caption{Comparison of uncertainty in QoI quantities from CPD calculations with equivalent values from SBM consistent parameter-sets across 15 DOE conditions.
Black dots indicate points generated from the CPD uncertainty analysis, red boxes correspond to QoI spaces used to judge consistency, nonblack dots are consistent SRWY model-form parameter-sets, and the color of the nonblack dots indicates the value of the high-temperature ultimate volatile yield $a$ for the consistent set.
The x axes are the time to half the ultimate volatile yield and y axes are the ultimate volatile yield value.}
\label{fig:Consistent_overlap}
\end{figure}

Now that a set of consistent points has been located, the SBM's performance at representing the CPD model can be further evaluated.
First, the consistent parameter-sets can be mapped into QoI space as shown in Figure~\ref{fig:Consistent_overlap}.
Clearly the red boxes, which were used as the QoI bounds within the consistency test, do not act as perfect representations of CPD's uncertainty, as is shown with the black dots taken directly from CPD's uncertainty quantification investigation.
With that caveat in mind, the red boxes do represent ranges of uncertainty believed to suffice in capturing the behaviors required of the SRWY model-form as specified by the application's requirements.
Again, information incorporated into the QoIs outside of CPD's uncertainty is evident in comparing the red boxes and black dots.

Within Figure~\ref{fig:Consistent_overlap} the value of parameter $a$ is shown as the color of the consistent points, or dots of colors other than black.
Interestingly, there is no observable correlation to $a$ values with the consistent point's positions within the 700 K DOE QoI spaces, but a linear correlation vertically across ultimate volatile yield is present in all other DOE conditions.
The triangular shape of the CPD uncertainty region at 700 K is also not captured.
The rectangular shape of CPD's QoI uncertainty is well replicated by the SRWY model-form consistent points for all higher temperatures.

A few additional features of the QoI spaces are captured poorly by the SRWY model-form.
For DOE conditions at 1,000 K the SRWY model-form was not able to be consistent for ultimate volatile yields below 0.45, while CPD yields spanned to near 0.4.
Similar issues are seen to lesser extents for higher temperature and lower heating-rate DOE conditions.  Back in Sec.~\ref{subsec:SBdecisions} it was decided that the SRWY model-form would not include the heating-rate as a functional input.
A consequence of this engineering decision can now be seen in the SRWY model-form's ability to match QoIs across the range of heating-rates for all temperatures above 1,000 K.
Consistent points are found in the initial portion of the temporal QoIs span for the 1E4 K/s DOE conditions, completely span 1E5 K/s DOE conditions, and overshoot much of CPD's uncertainty time-frame for 1E6 K/s DOE conditions.
This overshoot was permitted in order to allow the model to span greater amounts of the temporal QoIs within other DOE conditions.
As anticipated, the SWRY model-form performed best for the middle of the heating-rate range due to the yield model's parameter fitting method.
The fact that QoI uncertainty ranges were on the same order of magnitude as the heating rate's effect upon the ultimate volatile yield allowed the SWRY model-form to find consistency, but the need to increase the temporal bounds for the 1E6 K/s high-temperature DOE conditions demonstrates issues with this approximation.

\section{Model Credibility}
\label{sec:model_credibility}

Up to this point, SBM consistent parameter-sets proved basic adequacy conditions had been met.
Although consistent parameter-sets enable the SRWY model-form to meet specified QoI requirements, further analysis of the model results is required to judge the credibility of the model-form for use beyond this application.
While the model captures characteristics specified by the QoIs, does it appear physical in other attributes not considered with the QoIs?
Visualizing how the SRWY model-form traces compare with CPD traces allows further comparison of characteristics not quantified with the current QoI definitions, as was previously shown in Figure~\ref{fig:invert_energy}, and is a fundamental view of reaction model performance.
Additionally, analysis of model-form error or discrepancy between the SBM and CPD traces can provide evidence for the continued evolution of the model-form towards validity as well as qualified measures of credibility for the current model-form due to the credibility CPD possesses.
This type of analysis is similar to model validation, where model outputs are compared with experimental data that was not used for model calibration, except that no experimental data is currently used.

\begin{figure}[t!]
\centering
\includegraphics[width=0.9\columnwidth,keepaspectratio]{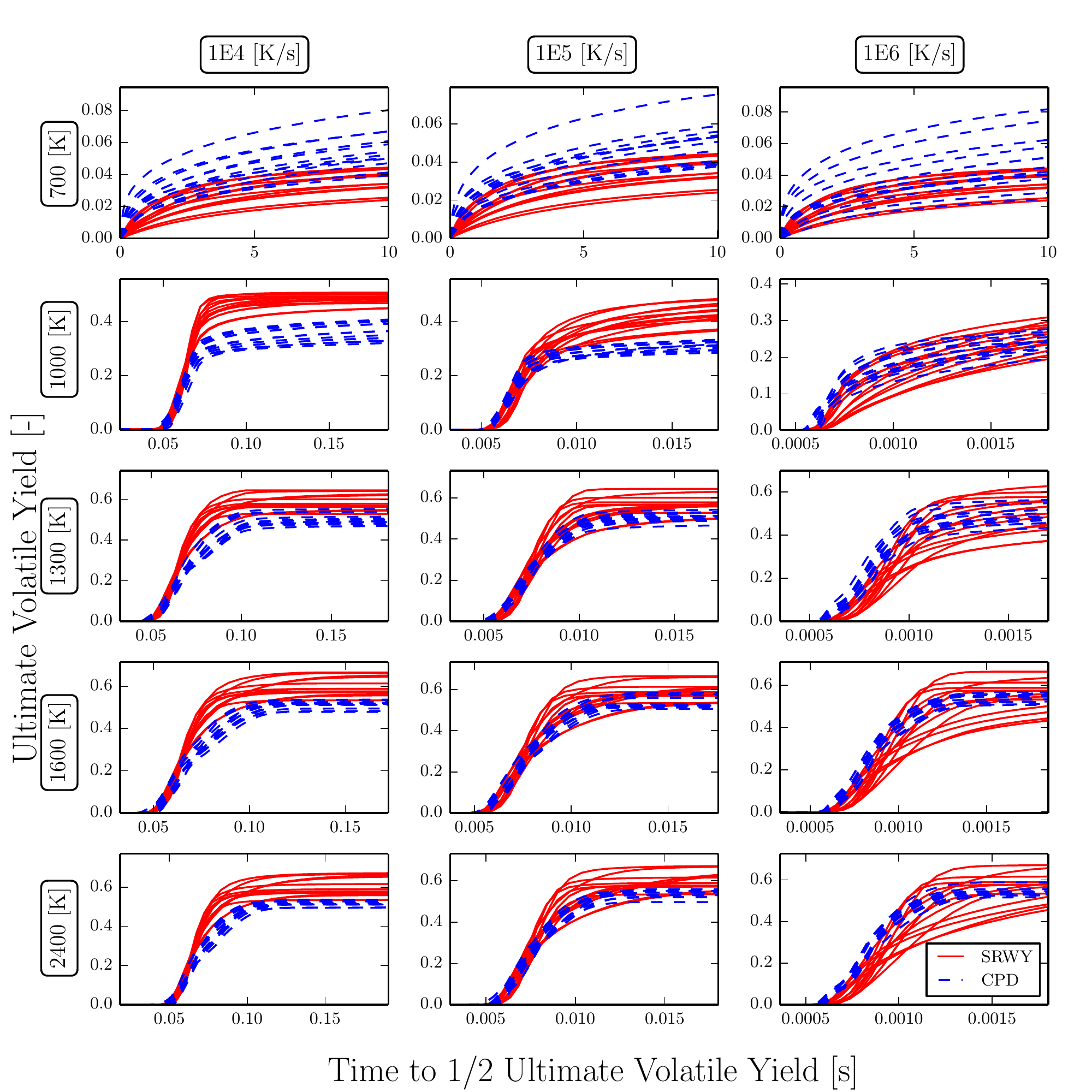}
\caption{Comparing traces generated with 12 randomly selected consistent SRWY model-form parameter-sets against 12 CPD traces created by randomly sampling the 13 uncertain parameters across 15 DOE conditions.
Only SRWY model-form parameter-sets with a non-inverted temperature distribution are used in this comparison.}
\label{fig:12traces}
\end{figure}

A comparison of 12 randomly selected consistent SRWY model-form traces and CDP traces is shown in Figure~\ref{fig:12traces}.
Only SRWY model-form traces corresponding to non-inverted activation temperature distributions are visualized due to those with inverted distributions previously being noted to have poor shape characteristics.
Kinetically, the SRWY model-form traces begin to display substantial amounts of devolatilization within the same time-frame as the CPD traces and all traces match the initial kinetic shapes satisfactorily.
Significant discrepancy then appears to occur within the two lower temperature DOE conditions during the later kinetic stages.
As could also be noted in Figure~\ref{fig:Consistent_overlap}, the SRWY traces do not reach as high of ultimate volatile yields as CPD traces at 700 K.
For the DOE conditions with 1,000 K hold-temperature and 1E4-1E5 K/s heating-rates, there is significant overshoot of the asymptote towards the ultimate volatile yield.
A systematic discrepancy can be noted in how the traces corresponding to 1E4-1E5 K/s heating-rates asymptote to the ultimate volatile yield.
The CPD traces have slower/gradual asymptotes, while the SRWY model-form's traces have sharper/abrupt asymptotes.
Greater variance in the SRWY model-form traces is also evident.  Such variance reflects that the current QoIs are not over-constraining the SRWY model-form.
The significant variation in the activation temperature distribution's standard deviation $\sigma_a$ spanning from zero to approximately ten percent ($\sim$ 4,000 K) of the activation temperature was expected to produce a wide assortment of kinetic shapes.
Such variation due to the activation temperature distribution was most evident in the 1E6 K/s traces.

Overall, the SRWY model-form's performance was deemed satisfactory for the application-scale's requirements.
While shortcomings of the SRWY model-form have been highlighted, its ability to meet strenuous demands for consistency with CPD characteristics across a wide range of system conditions is a strong statement towards its credibility.
The volatile yield traces produced by the SWRY model-form closely resemble the equivalent CPD traces.  
Basing our evaluation of the SWRY model-form on the application space demands for a scale-bridging approximation of CPD that can capture scale-specific physical descriptions, the SWRY model-form was deemed to meet the requirements.

\section{Model Refinement}
\label{sec:model_refinement}

Although time constraints necessitated the use of the model-form described thus far for the first year's simulations, utilizing discrepancies observed to motivate model-form refinement for year two's simulations was possible.
Propagation of knowledge gained is a fundamental piece of credible model development.
Due to the observed discrepancy correlated to low temperatures, factors effecting the SBM's performance in this condition region were reconsidered.

\begin{figure}[t!]
\centering
\includegraphics[width=0.4\columnwidth,keepaspectratio]{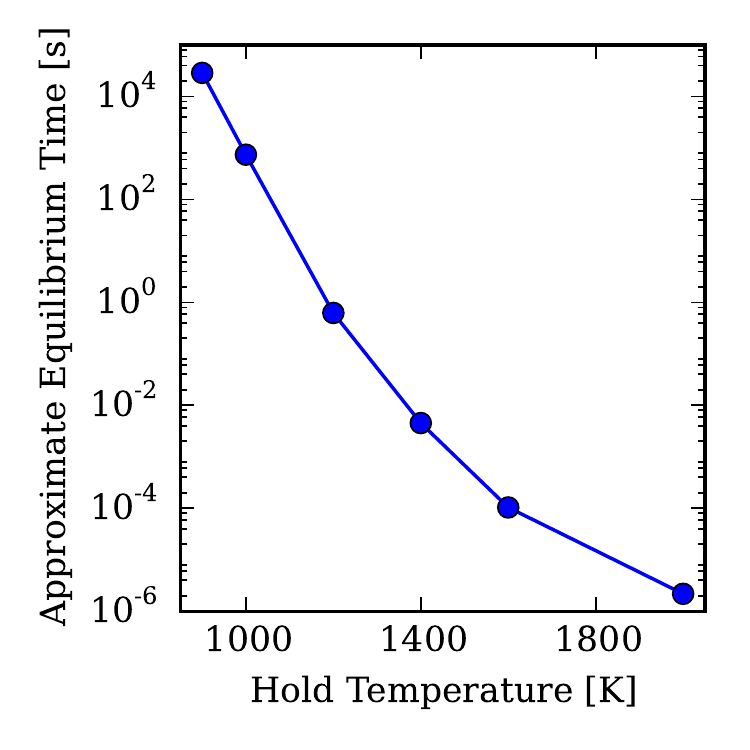}
\caption{Approximate thermodynamic equilibrium time for CPD traces across a range of hold-temperatures.
CPD traces were generated using nominal parameter values and a linear heating-rate of 1E9 K/s from 300 K to the respective hold-temperature.}
\label{fig:eqTimeScaling}
\end{figure}

Within Sec.~\ref{sec:heating_temp_effect} a hold-time of ten seconds was assumed to be the effective equilibrium time-scale for the applications of interest.
Figure~\ref{fig:ConstRtime} demonstrated that this assumption was not ideal for temperatures below 1,200 K even when the coal effectively experienced instantaneous heating.
While it is true that the coal will spend less than ten seconds in the boilers, this assumption reduced the driving force for reactions at lower temperatures.
The consequences of a reduced driving force were carried though the model development process, affecting the consistent spaces for the free parameters explored.
This is an illustrative example of the difficulty of performing calibration and quantifying model-form uncertainty simultaneously, also known as the identification problem \cite{Arendt2012}.
More realistic estimates of the time to reach equilibrium were investigated within Figure~\ref{fig:eqTimeScaling}, where nominal CPD parameter values were used.
Estimates of time to equilibrium  were determined by locating positions on yield traces where $(V_f - V)/V_f < 0.01$ or the yield was within 1\% of the ultimate yield for that hold-temperature.

As previously noted, ten seconds is a good approximation of an equilibrium hold-time above 1,200 K, but now it becomes evident that this is a poor assumption for the lower temperatures included in the DOE.
Data-points for temperatures below 900 K were not included in Figure~\ref{fig:eqTimeScaling} because the time to compute such data was prohibitive and the scaling behavior had already become evident.
Another important consideration is that this figure was created using a 1E9 K/s linear heat-up rate.
Slower heat-up rates would most strongly affect lower temperatures, lengthening the time to reach equilibrium.
Although the creation of an accurate yield model for low temperatures is out of scope for this study, increasing the hold-times for 500 K, 600 K, 700 K, 800 K and 900 K CPD runs to 1E5 seconds and 1,000 K to 1E4 seconds was viable.
The effect of incorporating this data into an improved yield model can then be used to determine if future model development iterations should focus upon this issue.

\begin{figure}[t!]
\centering
\includegraphics[width=0.5\columnwidth,keepaspectratio]{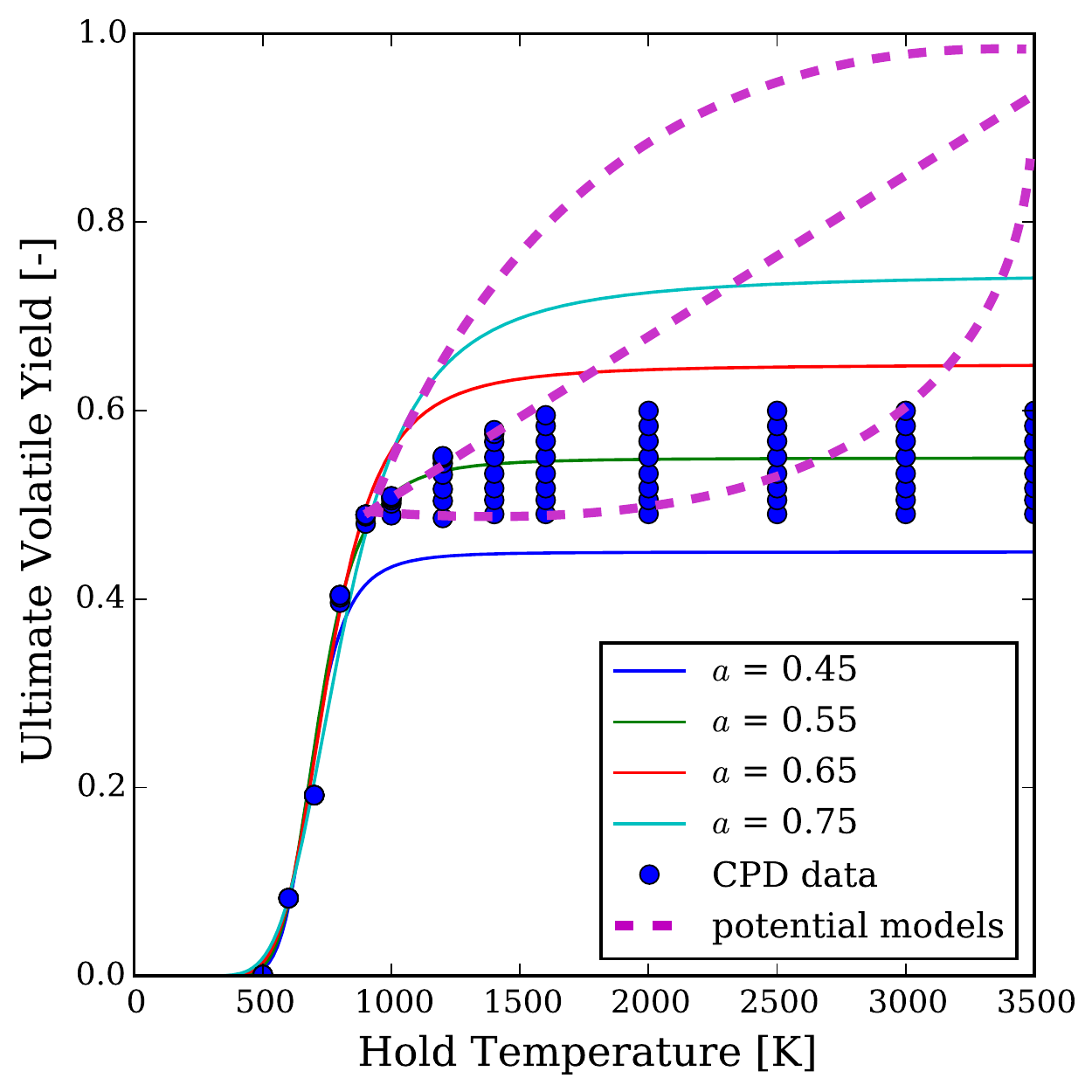}
\caption{Yield model equivalent to Figure~\ref{fig:yield_model} once the hold-times were increased for lower temperatures.
Solid lines are based upon Eq.~\eqref{eq:yield_model_eq}, where each line signifies a different $a$ value and the dots are CPD data with the heating-rates and hold-temperatures varied.
Additional purple dashed lines represent potential alternative forms of the yield model that could be implemented to account for uncertainty at higher temperatures.}
\label{fig:Yield_Model_Final_Form}
\end{figure}

Using the same methodology as was previously described within Sec.~\ref{sec:SBM}, a yield model was created from CPD data that spanned a range of heating-rates and hold-temperatures, but where the hold-times for lower temperatures were increased.
This adjusted yield model is shown in Figure~\ref{fig:Yield_Model_Final_Form}.
The parameters fit to the CPD data were b = 11.53, c = -9.122, d = 2.407, e = -0.7773, and 500 K was the devolatilization initiation-temperature.
The effect of longer hold-times becomes evident when the adjusted yield model is compared with Figure~\ref{fig:yield_model}.
The yield curve now reaches the high-temperature ultimate volatile yield close to 1,000 K instead of 1,250 K and there is less spread in the low temperature data points due to the heating-rate.

Also included in Figure~\ref{fig:Yield_Model_Final_Form} are purple dashed-lines indicating potential model-forms that could be explored in the future to account for uncertainty in the yields at higher temperatures.
Until data is available in such temperature regions, it will be difficult to compare potential model-forms and reach definitive conclusions.
Even if high-temperature experimental data for pure devolatilization does not become available in the near future, model-forms such as those suggested could be tested within multi-physics simulations against data-forms available for comparison at that scale, or top-down validation.

\begin{figure}[t!]
\centering
\includegraphics[width=0.9\columnwidth,keepaspectratio]{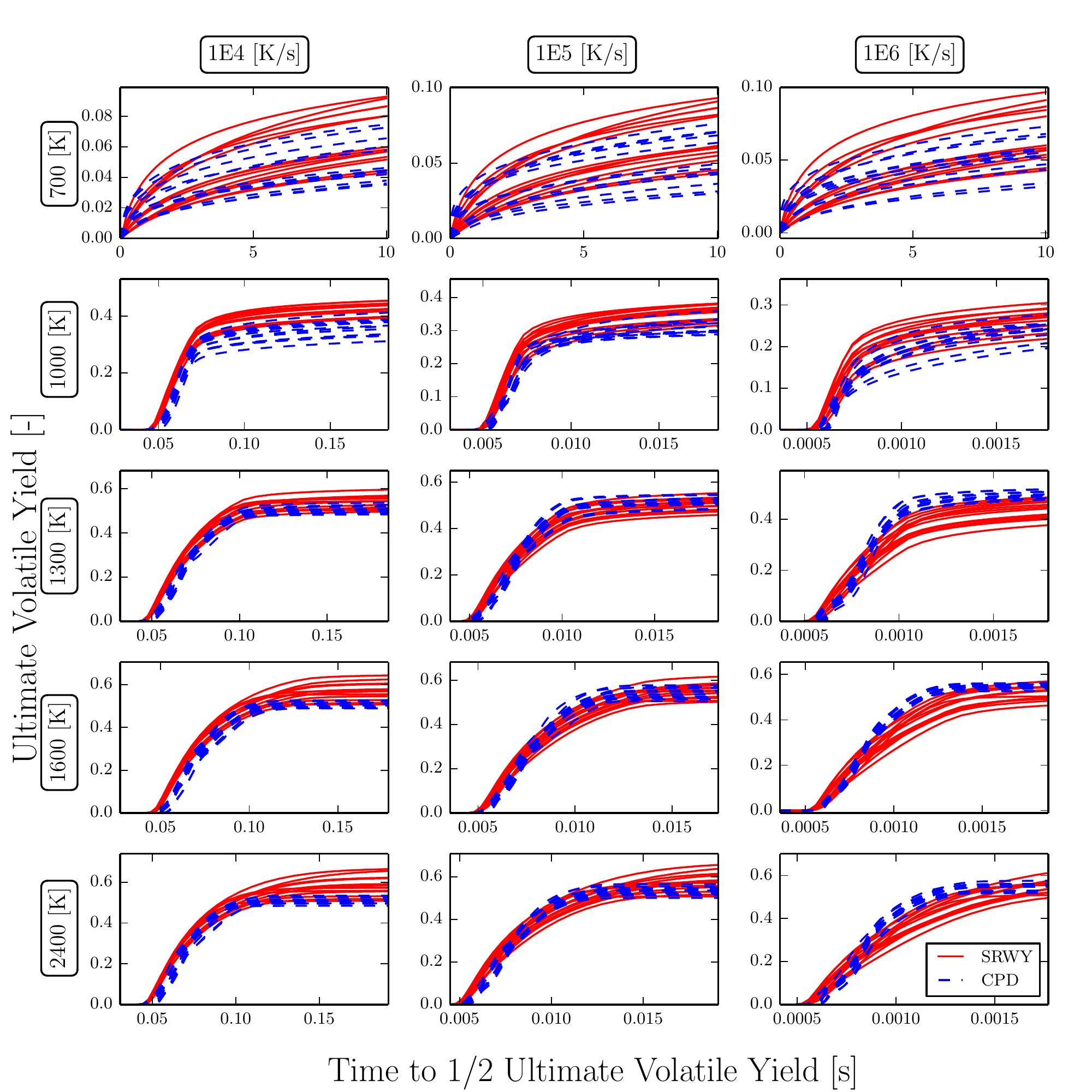}
\caption{A comparison of 12 traces generated with randomly selected consistent SRWY model-form parameter-sets against 12 CPD traces created by randomly sampling the 13 uncertain parameters across 15 DOE conditions.
The SRWY model-form used the yield model-form based upon longer times to reach equilibrium for lower temperatures.}
\label{fig:traces_cycle2}
\end{figure}

The updated yield model was used within Eq.~\eqref{eq:SRWY} and consistent parameter-sets were located in the manner previously described within Sec.~\ref{sec:consistency_eval}.
The impact of the improved yield model can be judge through a yield trace comparison, as shown in Figure~\ref{fig:traces_cycle2}.
Compared to Figure~\ref{fig:12traces}, improved performance in matching the low-temperature yield trends is evident.
Additionally, the traces demonstrated improved matching of CPD's asymptotic behavior across the higher temperature traces.
Improvement in the asymptotic behavior appears to be at least in part caused by the activation temperature distribution's standard deviation $\sigma_a$ now varying between 3,000 K and 8,000 K, where previously it ranged from 0 K to 5,000 K.
The distribution's width is still of similar breadth, but the updated yield model caused the entire distribution to shift towards larger values.
The previously noted variance in the appearance of the volatile yield trends has also been significantly reduced.
While again this is likely due to a combination of factors, the large reduction in widths of the consistent ranges for $A$ and $E$ are likely major contributors.
With the updated yield model, optimal parameter-set values were found to be $\log_{10}(A) = 8.499 \, s^{-1}$, $E = 14,380 \, K$, $\sigma_a = 4,719 \, K$, and $a = 0.565$.
These optimal values are similar to those previously found, when inverted energy distributions were not considered, except that the activation temperature distribution's value is approximately six times larger.

An additional attribute of the current DOE becomes evident once the implications of Figure~\ref{fig:eqTimeScaling} for high temperatures is considered within Figure~\ref{fig:traces_cycle2}.
For all three considered heating-rates, the traces above 1,000 K appear to be approximately the same.
For the fastest heating-rate considered, 1E6 K/s, it takes approximately 0.001 seconds to reach 1,000 K and the thermodynamic time-scale is over 100 seconds at that temperature.
But for 1,600 K the thermodynamic time-scale is approximately 1E-4 seconds, meaning that the traces are heating-rate limited and effectively are thermodynamic yield curves.
Evidently, future iterations of model-development could remove the 2,400 K DOE conditions and likely benefit from additional lower temperature DOE conditions or higher heating-rate conditions where the heating-rate effects could be further isolated from the kinetics.
Moving forward it is noted that the extended temporal range QoIs for higher temperatures and 1E6 K/s heating rates could be removed, due to consistency no longer necessitating this extension.

Clearly, the updated yield model positively impacted the SBM's credibility.
Given the performance noted across traces for all DOE conditions, there should be greater confidence in utilizing the SBM in interpolative applications and even slight extrapolations.
Throughout the developmental process of creating the current model-form, tasks that could be completed to increase the model's credibility for lower or higher temperature applications have been detected.
Lower temperature applications would benefit from additional refinement of the yield model through longer CPD runs.
Higher temperature uses could look to the creation of experimental data or validate inversely through comparing multi-physics simulations with experimental data available at the application scale.

\section{Conclusions}
\label{sec:conclusions}

The need for a computationally cheap function to capture scale-appropriate traits of a detailed physics model, which has been deemed to contain significant amounts of uncertainty, was the driving force behind this research.
Once the desired characteristics of the detailed model were determined and corresponding uncertain quantified, a SBM was created using a single reaction model with functional yield model and distributed activation energy.
This SBM contained four free-parameters, which were calibrated using consistency constraints against selected QoIs.
The QoIs were based upon capturing desired physics and were quantified by the uncertainty contained within the detailed model and additional insights.
Once consistent parameter-sets were located, the credibility of the SBM in representing the detailed model for the application-scale was evaluated.
Through visualizing the mapping of the consistent parameter-sets into the space of the QoIs, qualitatively comparing the characteristics of kinetic traces and considering cumulative distributions, the SBM was deemed to have performed satisfactorily for the stated application.
Then utilizing the discrepancies discovered throughout the model assessment, an improvement to the yield model was implemented and the gained performance demonstrated.

The model development and analysis demonstrated throughout this work were based upon one particular coal type, Utah Sufco bituminous.
The SWRY model-form can be applied to alternative coal types through repeating steps used during the model development process.
The most significant difference between coal types will come from fitting the yield model to coal specific CPD thermodynamic data.
Once alternative fitting parameters have been found for the yield model, consistency test can be rerun.
The parameter-space exploration should be expedited by utilizing consistent regions found for the current coal type as prior knowledge to base parameter bounds upon.

Like all engineering exercises, this process has the potential for continued refinement.  Incorporation of addition sources of data, especially experimental data, is an obvious next step.
High-temperature experimental-data could greatly reduce the uncertainty in the model-form for higher-temperature applications.
An alternative to experimental data could be a comparison against another detailed model such as FLASHCHAIN \cite{Niksa1991}, which could give further credibility to the SBM within its current areas of application.
Continued alteration to the form of the yield model is an additional avenue for simple improvement that has been demonstrated to positively impact the model's performance and which could include model-form comparisons.
Reconsidering the DOE design is another easy alteration to incorporate in order to gain additional insight efficiently.
As previously stated, QoI definitions are subjective, so further exploration might lead to superior model performance.
Balancing the cost-benefit of such refinements is a research area in its own right.

\section*{Acknowledgements}
This material is based upon work supported by the Department of Energy, National Nuclear Security Administration, under Award Number(s) DE-NA0002375.

\bibliographystyle{elsarticle-num}
\bibliography{devol_doc.bib}

\end{document}